\UseRawInputEncoding
\documentclass[%
superscriptaddress,
 amsmath,amssymb,
 aps,
 pra,
]{revtex4-2}

\usepackage{graphicx}% Include figure files
\usepackage{dcolumn}% Align table columns on decimal point
\usepackage{bm}% bold math
\usepackage{float}
\usepackage{color}

\begin{document}

\preprint{APS/123-QED}

\title{Strongly Electromechanical Coupled Phononic Waveguides in Aluminum Scandium Nitride on Silicon Carbide}

\author{Yuanchen Deng}
    \affiliation{Center for Integrated Nanotechnologies, Sandia National Laboratories, Albuquerque, NM, USA}%

\author{Dalton Anderson}%
    \affiliation{College of Optical Sciences, University of Arizona, Tucson, AZ, USA}%

\author{Xingyu Du}
    \affiliation{Department of Electrical and Systems Engineering, University of Pennsylvania, Philadelphia, Pennsylvania 19104, USA}%
    
\author{Will Roberts}
    \affiliation{College of Optical Sciences, University of Arizona, Tucson, AZ, USA}%

\author{Michael Miller}
    \affiliation{Microsystems Engineering, Science, and Applications, Sandia National Laboratories, Albuquerque, NM, USA}%

\author{Brandon Smith}
    \affiliation{Microsystems Engineering, Science, and Applications, Sandia National Laboratories, Albuquerque, NM, USA}%
\author{Lisa Hackett}
    \affiliation{Microsystems Engineering, Science, and Applications, Sandia National Laboratories, Albuquerque, NM, USA}%
\author{Troy Olsson}
    \affiliation{Department of Electrical and Systems Engineering, University of Pennsylvania, Philadelphia, Pennsylvania 19104, USA}%
\author{Matt Eichenfield}
    \email{eichenfield@arizona.edu}
    \affiliation{College of Optical Sciences, University of Arizona, Tucson, AZ, USA}%
    \affiliation{Microsystems Engineering, Science, and Applications, Sandia National Laboratories, Albuquerque, NM, USA}%

\date{\today}% It is always \today, today,

\begin{abstract}

Guided phonons have become an increasingly important platform for classical and quantum information processing. While conventional surface acoustic wave systems are typically only guided in the vertical direction, two-dimensionally confined waveguide systems offer significant advantages in terms of density of phononic circuit components and much higher intensities of strain and piezoelectric fields, which make them promising candidates for advancing acoustoelectric and quantum phononic applications. One such material system for generating and guiding phonons at gigahertz frequencies is AlScN on SiC, which can be synthesized by sputter depositing AlScN directly onto SiC wafers. The AlScN on SiC platform allows for tightly vertically confined acoustic modes with high electromechanical coupling, high speed of sound, and simple fabrication of strip and rib waveguides. Until now, this system has only been studied as a slab waveguide platform, i.e., without any lateral waveguiding. Here, we present a 2D-confined phononic waveguide architecture in AlScN on SiC with strongly electromechanically coupled modes that could serve as a platform for phononic routing, power-efficient active and nonlinear phononic devices such as amplifiers, mixers, and oscillators, as well as for interacting with quantum systems such as vacancy centers, charge carriers, photons, and spins. We study two distinct gigahertz frequency waveguide mode families using impedance-matched interdigital transducers and characterize their electromechanical coupling and propagation losses. Additionally, we analyze how these waveguides could interact with various important quantum and classical systems that can be either embedded in SiC or heterogeneously integrated on the surface.

%two waveguide mode families by fabricating acoustic waveguide delay lines with varying lengths. Additionally, we have theoretically calculated the AE power dissipation and nonlinearities achievable in this platform, underscoring its strong potential for applications in both classical and quantum information processing systems.

\end{abstract}

\maketitle

%\tableofcontents

\section{Introduction}

Piezoelectric phononic systems are rapidly emerging as important components in both classical and quantum information processing systems. In classical information processing, these systems have demonstrated exceptional performance in radio-frequency (RF) signal processing components such as filters, delay lines, resonators\cite{DavidMorgan2007SurfaceFilters}, and more recently, high-gain and low-noise amplifiers\cite{Hackett2023Non-reciprocalOperation,Mansoorzare2022MicromachinedProcessing,Mansoorzare2022Ultra-WidebandTransducers,Hackett2024S-bandArchitecture} and high-efficiency wave mixers\cite{Hackett2024GiantHeterostructures,Mansoorzare2023Acoustoelectric-DrivenWaveguides}. Beyond classical applications, phononic systems are also showing significant promise in quantum information science. Specifically, they are being explored for strain-tuning of vacancy centers\cite{Falk2014ElectricallyCenters,Udvarhelyi2018iAbCarbide,Whiteley2019SpinphononAcoustics,Maity2020CoherentDiamond,Dietz2023Spin-acousticCarbide,Kuruma2024ControllingCrystals}, mechanical spin-photon control\cite{Clark2024NanoelectromechanicalChip}, linear mechanical quantum computing\cite{Qiao2023SplittingComputing}, single phonon detection, quantum memories \cite{Boisen2018SuperconductingStorage,Chu2018CreationResonator,Hann2019Hardware-EfficientSystems,Yang2025FluorescenceMicrocavity}, as well as for facilitating phonon coupling to superconducting qubits\cite{Dong2012ContinuousNetworks,Golter2016CouplingState,Chu2017QuantumQubits,Manenti2017CircuitWaves,Arrangoiz-Arriola2018CouplingCavity,Moores2018CavityRegime,Lemonde2018PhononWaveguides,Satzinger2018QuantumPhonons,Cleland2019MechanicalMachines,Bienfait2019Phonon-mediatedEntanglement}, positioning them for important roles in quantum computing architectures.

Despite these advances, existing piezoelectric phononic system designs exhibit several limitations that hinder their overall performance. Typical designs of the phononic devices are shown in Fig. 1(a). Conventional designs adopt slab phononic modes in the piezoelectric layer or the non-piezoelectric substrate and exhibit low phonon intensity and an insufficient piezoelectric field in comparison with the waveguided phononic system. Without 2D-confined phonon propagation in piezoelectric materials, these systems are constrained not only in the efficiency and nonlinearity of acoustoelectronic (AE) devices, but also the phonon coupling strength with quantum components such as color centers and superconducting qubits. Phononic waveguide platforms, such as unreleased rib\cite{Pant2011On-chipScattering,Fu2019PhononicPhonons,Wang2020High-acoustic-index-contrastModeling,Mayor2021GigahertzSapphire,Bicer2022GalliumDevices,Guida2024DesignCircuits}, suspended\cite{Hatanaka2014PhononCircuits,VanLaer2018ElectricalCircuit,Lemonde2018PhononWaveguides,Liu2019ElectromechanicalWaveguides,Romero2019PropagationWaveguide,Sarabalis2020S-bandNiobate,Dahmani2020PiezoelectricWaveguide,Lu2020GHzLines,Kurosu2020MechanicalWaveguide,Mansoorzare2022MicromachinedProcessing,Ding2024IntegratedDiamond}, and overlay slot waveguides\cite{Feng2023GigahertzWaveguides}, as well as phononic crystals\cite{Sun2006PropagationMethod,Vasseur2007WaveguidingPlates,Mohammadi2009High-QSlab,Khelif2010AcousticSlabs,Balram2016CoherentCircuits,Patel2018Single-ModeWire,Pourabolghasem2018WaveguidingSlabs,Wang2019HexagonalWaveguides}, have been developed to mitigate some of these challenges. Among these designs, unreleased rib waveguides have provided significant advantages in heat dissipation and mechanical robustness while maintaining outstanding scalability. In unreleased rib waveguides platforms, phonons can interact not only with the color centers and carriers that are embedded within the substrate but also with quantum elements such as superconducting resonators and qubits which are usually placed on top of the substrate as shown in Fig.1 (b). However, these platforms often suffer from drawbacks such as low electromechanical coupling coefficients, poor thermal conductivity and limited integration with semiconductor technologies. Moreover, the common phononic waveguide modes in these platforms usually generate weak strains in the substrate and cannot transform into substrate surface phonons efficiently, which further constrain their potential especially for quantum phononic systems. 

To address these challenges, we propose a novel phononic waveguide architecture constructed with aluminum scandium nitride (AlScN) thin films on silicon carbide (SiC). The introduction of scandium into aluminum nitride films, forming an AlScN alloyed film, has shown significant increase of the piezoelectric constant compared to bare aluminum nitride\cite{Chen2023Scandium-DopedApplications}. This results in acoustic modes having larger electromechanical coupling coefficients\cite{Hackett2023AluminumFrequencies,Du2024NearSiC}. Our phononic waveguide platform exhibits several performance benefits compared to previously reported non-suspended waveguide platforms. 1) The phononic mode family in our waveguides exhibits strong electromechanical coupling, second only to non-suspended LiNbO$_3$ phononic waveguide platforms\cite{Du2024NearSiC,Mayor2021GigahertzSapphire}. 2)Due to the high velocity contrast between AlScN and SiC, resulting in a shorter evanescent length, the acoustic mode families display greater strains and displacements closer to the interface of AlScN and SiC, as shown in Fig. 1(c). 3) One of the mode families, the Sezawa mode shown in Fig. 1(d), could easily transfer into SiC surface wave phonons, which facilitates coupling to superconducting resonators and qubits on bare SiC surfaces. 4) The Sezawa mode family possesses a higher speed of sound, which is approximately twice the velocity of the waveguide mode in solidly anchored LiNbO$_3$ platforms\cite{Du2024NearSiC,Mayor2021GigahertzSapphire}, therefore enabling higher operating frequencies with the same lithography resolution. 5) AlScN films can be sputtered directly on SiC with good lattice matching with graded Sc concentration interface films\cite{Hackett2023AluminumFrequencies,Du2024NearSiC}, offering orders of magnitude higher cross-wafer uniformity than thin-film LiNbO$_3$\cite{Luke2020Wafer-scaleCircuits}, which is critical for large phononic devices and circuits where the cross-wafer uniformity is vital for ensuring uniform dispersion for the system. 6) SiC has an outstanding thermal conductivity enabling efficient heat dissipation in densely integrated on-chip devices, especially acoustoelectric devices which actively dissipate power\cite{Hackett2023Non-reciprocalOperation, Hackett2024S-bandArchitecture,Hackett2024GiantHeterostructures}. Additionally, SiC has shown great potential as a platform for quantum applications such as strain-coupled vacancy centers \cite{Falk2014ElectricallyCenters,Castelletto2020SiliconApplications} and superconducting resonators\cite{Buijtendorp2022HydrogenatedCircuits}. These advantages make this platform highly suitable for both classical and quantum information processing applications.

\begin{figure}
\includegraphics[width=16cm]{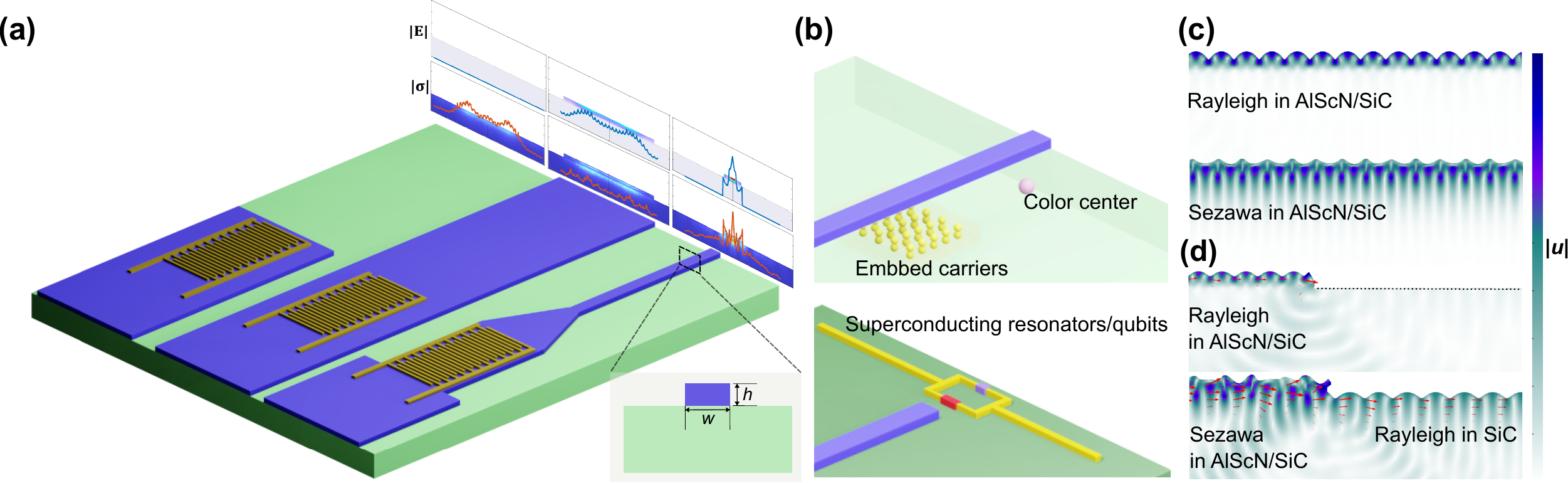}
\renewcommand{\figurename}{Fig.}
\caption{\label{fig:1} (a) Schematic of three different types of phononic devices. The top right inset figures at the edge of devices are normalized electric fields (top) and stress per phonon (bottom) in the corresponding phononic device. The bottom right inset plot is the cross-section of an AlScN phononic waveguide on a SiC substrate. (b) Schematic of integrating the phononic waveguides with different quantum elements.
(c) Simulated displacement profiles of the Rayleigh and Sezawa modes for continuous AlScN on SiC system. 
(d) Simulated displacement profiles of Rayleigh and Sezawa phonons for terminated AlScN on SiC, highlighting the Sezawa modes ability to couple to a SiC SAW. Red arrows represent the mechanical energy flux.
}
\end{figure}

In this study, we demonstrate solidly mounted integrated phononic waveguides in a 1 $\mu$m AlScN film with $42\%$ scandium concentration on a 4H-SiC substrate. At a wavelength $\lambda = 1.6$ $\mu$m, the AlScN waveguide hosts two phononic mode families with approximate velocities of $4780$ m/s and $6480$ m/s. These two modes are a Rayleigh-like mode at 2.95GHz and a Sezawa-like mode at 4.05GHz. Both phononic modes are excited by impedance-matched interdigital transducers and coupled into the waveguides through angled tapers. We have fabricated and subsequently tested the phonon transmission in straight waveguides with interdigital transducers at each end to characterize both the phonon propagation losses and the taper insertion loss. The AlScN waveguides demonstrate a propagation loss of the Sezawa mode that is approximately twice that observed in slab phononic delay lines. To highlight the potential of this platform, we have evaluated the enhancement in acoustic energy (AE) power dissipation and nonlinearities in phononic waveguides achieved using an indium gallium arsenide (InGaAs)/aluminum scandium nitride (AlScN)/silicon carbide (SiC) heterostructure. The reduced power dissipation and improved AE nonlinearity underscore the AlScN/SiC structure's effectiveness as a promising platform for AE phononic waveguides. Additionally, we calculated the spin-strain coupling strength $\Omega_m$ based on a previous study of acoustically straining coupling of the divacancy centers in 4H-SiC, showing a significant advantage of the Sezawa mode in the interaction with embedded spins in SiC.

\section{Design and fabrication}

For ease of device design and fabrication, we selected an AlScN film thickness of $h=1$ $\mu$m that is similar to our previous work on acoustic-electric (AE) amplifiers using AlScN/SiC \cite{Hackett2024S-bandArchitecture}. In \cite{Du2024NearSiC}, the maximum electromechanical coupling coefficient $k^2$ for the Sezawa mode was experimentally observed at a thickness-to-wavelength ratio between 0.56 and 0.68. This corresponds to a wavelength range of 1.47 to 1.69 $\mu$m. In this work, we opted for a wavelength of $\lambda = 1.6$ $\mu$m. We conducted a finite-element eigenmode analysis of AlScN waveguide sections with varying widths $w$, with the dispersion diagram of $w = 3\lambda$ presented in Fig. 2(a). Two phononic waveguide modes were identified: a Sezawa-like mode(Fig. 2(b)) with a velocity of $6480$ m/s and a Rayleigh-like (Fig. 2(c)) mode with a velocity of $4780$ m/s. The Sezawa-like mode exhibits more strain near the interface between the piezoelectric film and the substrate, as opposed to the Rayleigh-like mode that has a strain maximum at the structure's top surface. The electromechanical coupling coefficients of these two modes is given by $k^2 = 2(v_o-v_s)/v_o$, where $v_o$ and $v_s$ are mode velocities under open and shorted boundary conditions. The simulations give an electromechanical coupling coefficient of $k^2 = 4.27\% $ for the Sezawa-like mode and $k^2 = 0.2\%$ for the Rayleigh-like mode. 

\begin{figure}[htbp]
\includegraphics[width=12cm]{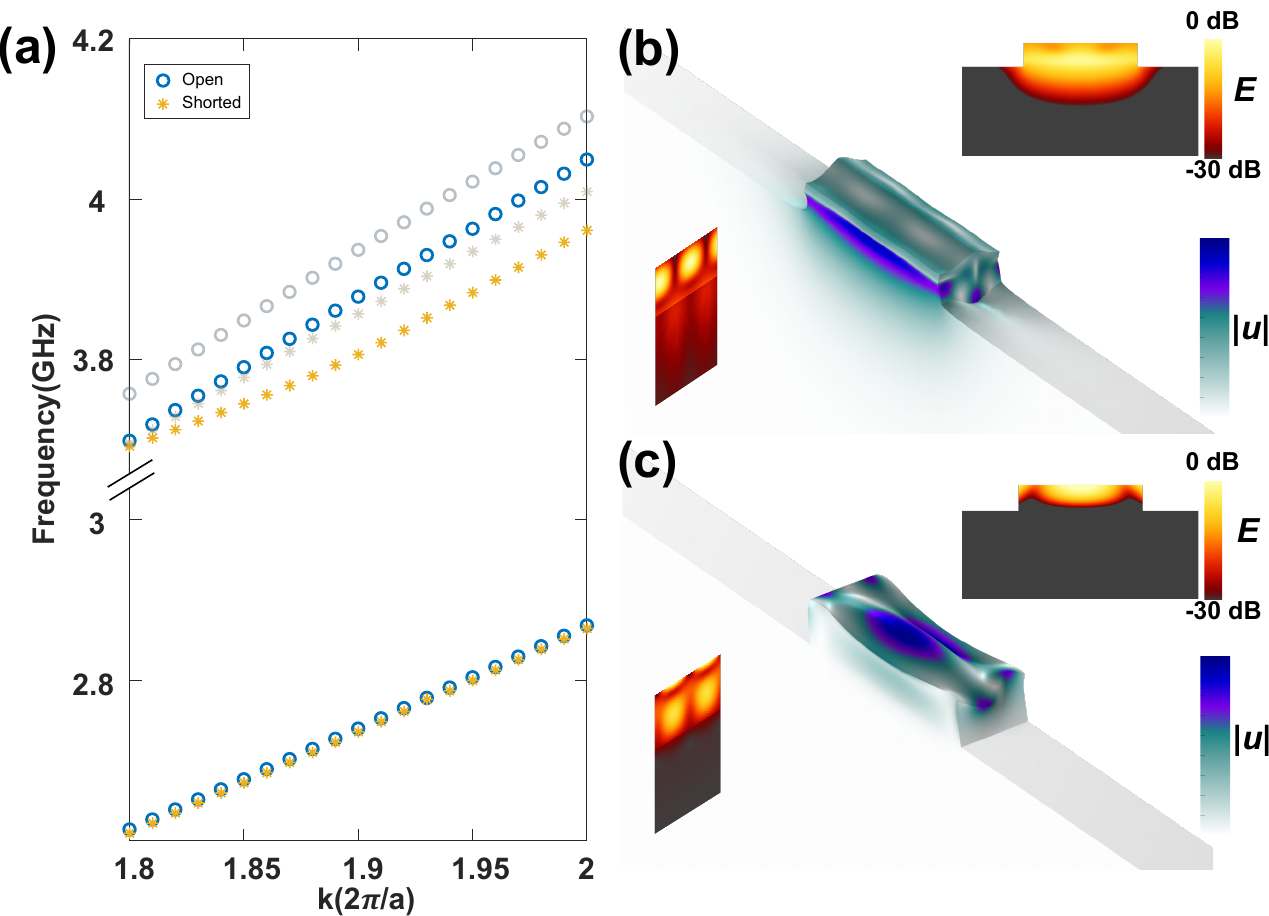}
\renewcommand{\figurename}{Fig.}
\caption{\label{fig:2} (a) The dispersion diagram of the phononic waveguide modes ($w=3\lambda$) under open and shorted boundary conditions, represented by circles and asterisks correspondingly. The grey color represents another Sezawa-like mode around 4GHz.(See Supplementary Material section B). (b), (c) Total displacement profiles of the Sezawa-like mode (b) and the Rayleigh-like mode (c) in acoustic waveguides. The top-right inset heat maps are cross-sectional acoustic energy distributions of the corresponding waveguide mode. The bottom-left heat maps are strain fields along the propagation direction. }
\end{figure}

To excite the surface phonons in AlScN, we designed interdigital transducers (IDTs) that match the impedance of the RF transmission line (50$\Omega$). The aperture width of the IDT is much larger than the waveguide width to guarantee a reasonable bandwidth. We analyzed the impedance of the IDT in COMSOL Multiphysics with various aperture widths and numbers of electrode pairs (see Supplementary Materials section C). We selected the Sezawa-like mode as the optimization target to take advantage of its higher electromechanical coupling coefficient. The IDTs are then patterned by electron-beam lithography using a bilayer polymethyl methacrylate (PMMA) resist. A 10 nm layer of titanium followed by a 90 nm layer of aluminum is then deposited, and a lift-off process is carried out in N-Methylpyrrolidone (NMP). As shown in Fig. 3(a), the fabricated IDT consists of 40 pairs of metal electrodes and has an aperture width of $123$ $\mu$m. The RF power reflection $|S_{11}|$, shown in Fig. 3(b), displays two significant dips corresponding to the Rayleigh-like mode and the Sezawa-like mode. The bandwidth of the Rayleigh-like mode is noticeably narrower than that of the Sezawa-like mode, likely due to a weaker electromechanical coupling coefficient and stronger acoustic reflections of the IDT electrodes. To further experimentally characterized the electromechanical coupling coefficient of the waveguide modes, we first extract the total impedance of the IDT $Z$ with $Z = 50 (1+S_{11})/(1-S_{11})$. The electromechanical admittance $Y = G+iB$ of the IDT can be obtained by $Y = Z^{-1}$ after the resistance and inductance within the measurement circuit are removed from the impedance by adopting the Mason equivalent circuit model (see Supplementary Material section D). The electromechanical conductance $G$ and susceptance $B$ are plotted in Fig. 3(c), showing aligned resonance peaks with the $S_{11}$ spectrum. The electromechanical coupling coefficients of the waveguide modes were extracted by $k^2 = \frac{\pi}{4}\frac{1}{\omega_0^2C_T}\int{G(\omega)d\omega}$, where $C_T$ is the capacitance of the transducer and $\omega_0$ is the angular resonance frequency of the waveguide modes\cite{Hackett2023AluminumFrequencies}. By numerically integrating the $G$, we obtained $k^2 = 6.08\pm1.2\% $ for the Sezawa-like mode and $k^2 = 0.76\pm0.18\% $ for the Rayleigh-like mode, which agree with the results from 2D cross-sectional finite-element simulation results. (See Supplementary Materials Section E). 

\begin{figure}[htbp]
\includegraphics[width=12cm]{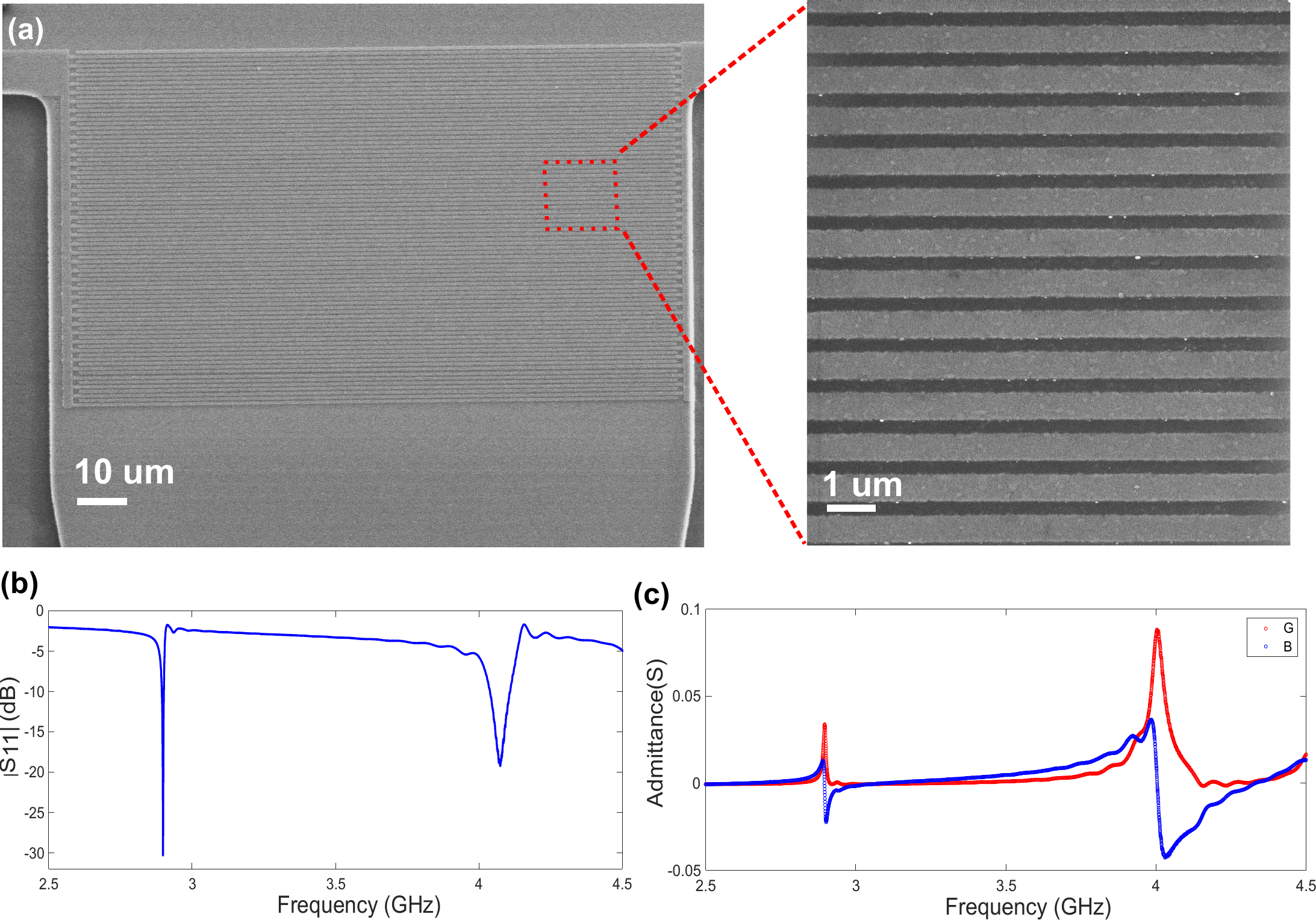}
\renewcommand{\figurename}{Fig.}
\caption{\label{fig:3}(a)Scanning Electron Microscopy (SEM) images of an IDT for launching surface acoustic waves into a waveguide taper. (b) RF power reflection $|S_{11}|$ of the IDT in (a).  (c) Real(red) and imaginary(blue) part of the electromechanical admittance of the IDT in (a).}
\end{figure}

%It is noted that the IDT has a metalization ratio greater than 0.5, i.e. a finger width larger than $\lambda/4$, due to fabrication imperfections from variations in Ebeam lithography. \YQ{While this could lead to a higher capacitance $C_T$ and a reduced bandwidth, it does not significantly affected the phonon excitation efficiency (see Supplementary Material section E)}.

Next, we used a wet chemical etching process, as shown in Fig. 4, to fabricate slab and straight waveguide phononic delay lines on an AlScN/SiC chip (Fig. 5(a) and (b)). First, a 300 $\mu$m layer of silicon dioxide (SiO$_2$) is grown on the AlScN/SiC chip using a plasma-enhanced chemical vapor deposition (PECVD) process. Next, a photoresist (PR) mask is patterned onto the SiO$_2$ through optical lithography. The SiO$_2$ is then patterned as a hard mask via an inductively coupled plasma (ICP) dry etching process using a CF$_4$/O$_2$/Ar plasma. After the photoresist is removed, the chip undergoes a through-etch of the 1 $\mu$m AlScN film in a 25\% tetramethylammonium hydroxide (TMAH) solution at 80$^\circ$C for 25 minutes. Finally, the SiO$_2$ hard mask is removed by dipping it in a 5\% hydrofluoric acid (HF) solution for 3 minutes. Two identical IDTs are patterned with electron beam lithography and deposited at both ends of these devices to excite and receive surface acoustic waves. In straight waveguide delay lines, the phonons are coupled into the waveguide via an angled taper with a length of 320 $\mu$m. The phonon propagation length $L$ varies from 200 $\mu$m to 2000 $\mu$m in both types of delay lines, excluding the lengths of both tapered regions.

\begin{figure}[htbp]
\includegraphics[width=16cm]{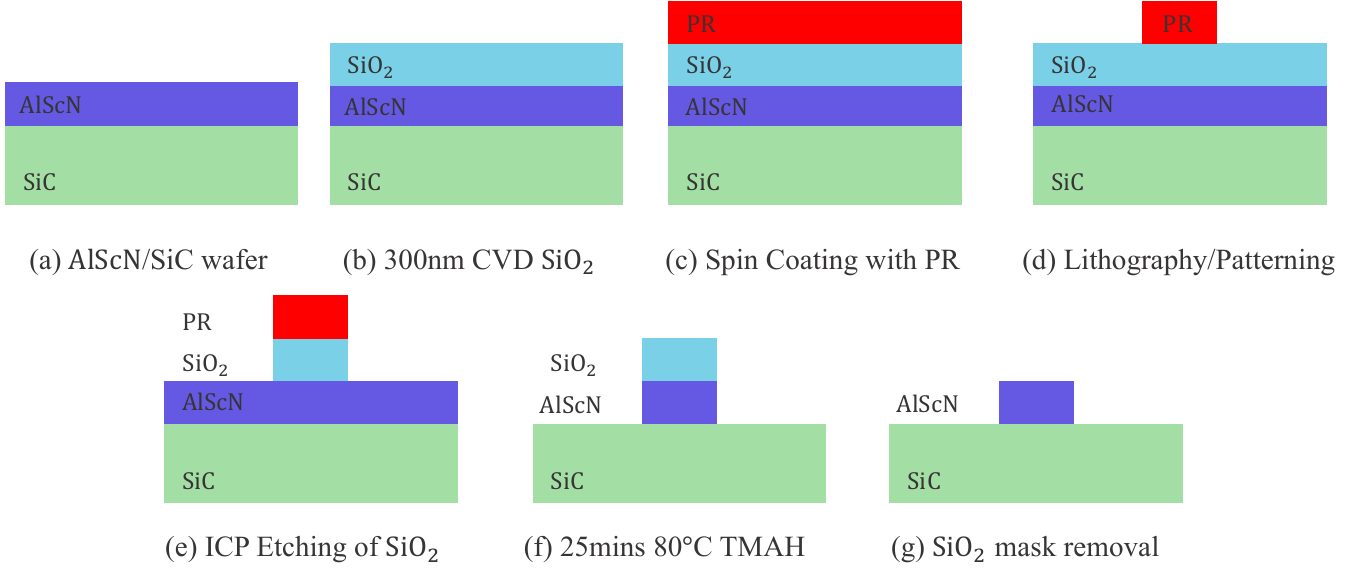}
\renewcommand{\figurename}{Fig.}
\caption{\label{fig:4} Fabrication steps of AlScN waveguides on SiC.}
\end{figure}

\section{Loss characterization}
In this section, we measured the transmission in all delay lines from 2GHz to 5GHz. Microwave scattering parameters $S_{ij}$ transmission of the delay lines are taken with Keysight KT-P9374A vector network analyzer (VNA) with FormFactor ACP40-GSG-150 GSG probes contacting the IDTs. S-parameters for Rayleigh (2.95Ghz) and Sezawa (4.05Ghz) modes are collected for 72 different devices varying the propagation lengths of the waveguide and slab devices. Devices with poor performance due to fabrication imperfections resulting in poor IDT responses were omitted. Fig. 5(c) shows $|S_{21}|$ spectra for examples of two different delay lines: a slab delay line and a straight waveguide delay line with $L=320$ $\mu$m. We observed that the transmission peaks for the Rayleigh-like (2.95 GHz) and the Sezawa-like (4.05GHz) modes in straight exhibit noticeably lower amplitude than slab delay lines, indicating additional transmission losses in these devices.

For an acoustic delay line, the total transmission loss can be written as $\mathcal{L}_{total}=\alpha L +2 \mathcal{L}_{taper} + \mathcal{L}_{i}$, where $\mathcal{L}_{taper}$ is the loss from each tapering region ($\mathcal{L}_{taper} = 0$ for slab devices), $\mathcal{L}_{i}$ is the total IDT insertion loss, $\alpha$ is the propagation loss per unit length of the waveguide, and $L$ is the length of propagation. To characterize the propagation losses, we linearly fit the maximum $|S_{21}|$ vs propagation length of the Sezawa-like mode for both types of devices, as shown in Fig. 5(d). The linear fit of Sezawa slab delay lines gives an intrinsic propagation loss $\alpha_{slab}=5.3\pm 0.2$ dB/mm, while the Sezawa straight waveguide delay lines loss is measured as $\alpha_{waveguide} = 10.7\pm 1.7$ dB/mm. The higher propagation loss in the straight waveguide, compared to the slab, is likely due to propagation losses caused by rough waveguide edges from the wet etching process (see Supplementary Material section F). Additionally, assuming equal total IDT loss ($\mathcal{L}_{i}$), the difference in the linear fit intercept between the slab and straight waveguide delay lines represents the insertion loss from the two identical tapered regions and is measured as $2 \mathcal{L}_{taper} = 19.8\pm8.3$ dB.  This suggests that a single taper contributes $\mathcal{L}_{taper} = 9.9\pm 4.1$ dB of loss, which can be further optimized by reducing the tapering angle or adopting a focused IDT design\cite{Guida2024DesignCircuits}. A similar analysis for the Rayleigh-like mode, which has a lower $k^2$ and a narrower bandwidth, is discussed in Supplementary Material section G.

\begin{figure}[htbp]
\includegraphics[width=16cm]{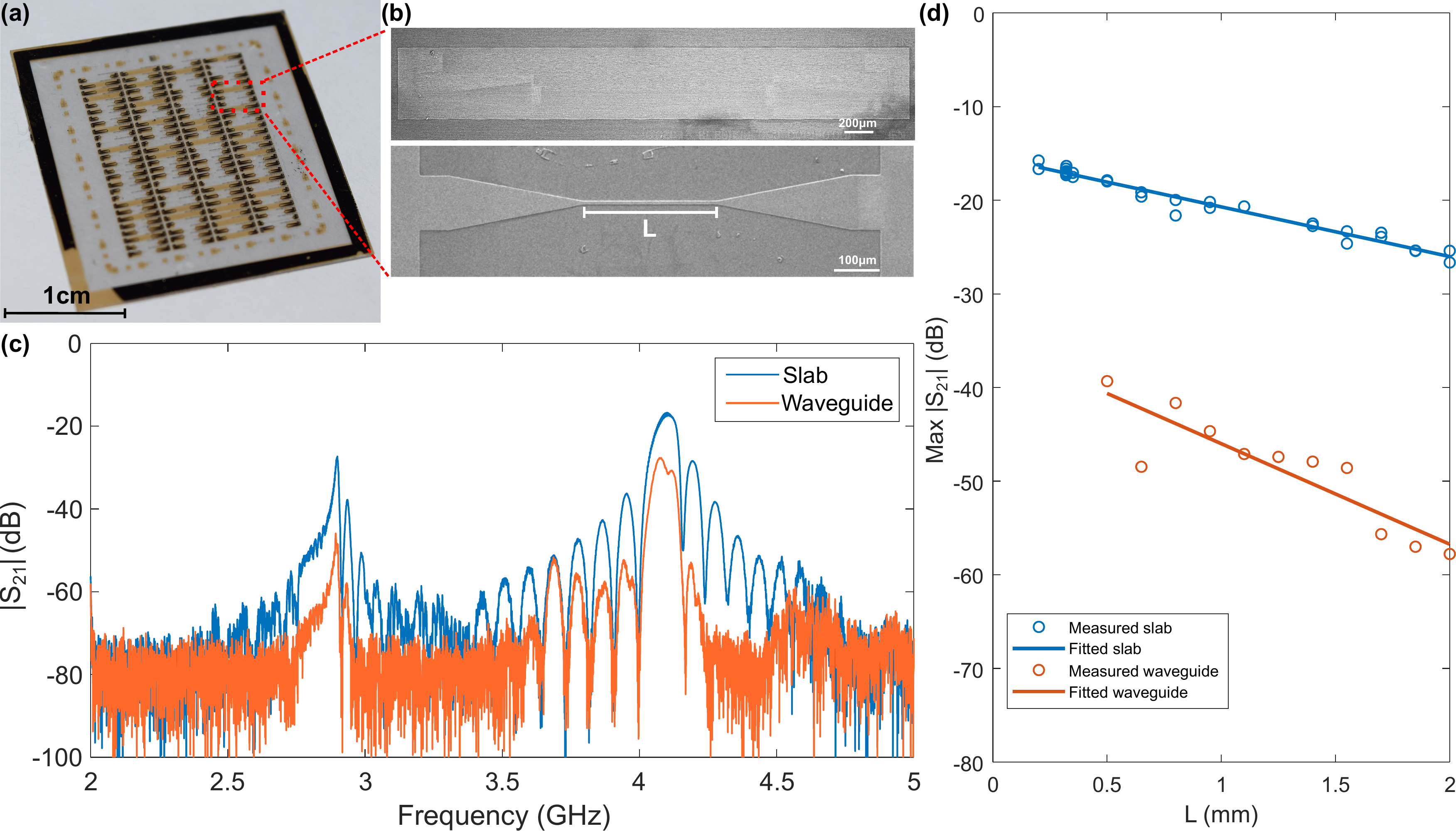}
\renewcommand{\figurename}{Fig.}
\caption{\label{fig:5} (a) The photograph of the fabricated AlScN/SiC chip with phononic delay lines. (b) SEM images of the slab (top) and waveguides (bottom) respectively. (c) Example $|S_{21}|$ transmission data for slab (blue) and waveguide (orange) delay lines with $L=320$ $\mu$m. (d) Maximum $|S_{21}|$ transmission for the Sezawa-like mode in varied length devices with linear fits to determine propagation and insertion losses.}
\end{figure}

%For the S-bend delay lines, the linear fit yielded $|S_{21}|=-(10.0\pm 3.7)L-(42.75\pm12.47)$ dB, with a slope similar to that of the straight waveguide delay lines.

%\DA{The extra propagation of the Euler bend, $\mathcal{L}_{bend} = \langle \alpha_{bend} \rangle L_{bend} \approx 7.52$ dB, can be extracted from the increase in the intercept parameter as all S-bend delay lines have the same Euler bend($L_{bend} = 345 \mu$m). From this, the average Euler bend propagation loss can be determined $\langle \alpha_{bend} \rangle \approx 21.7$ dB/mm.}

\section{Enhanced Acoustoelectric Efficiency}
The AlScN/SiC platform has demonstrated exceptional performance for AE amplifiers\cite{Hackett2024S-bandArchitecture}. Here we theoretically evaluate the improvement brought by a phononic waveguide design in both AE amplifiers and AE wave mixers. As shown in Fig. 6(a), the waveguide AE device features a semiconductor layer of the same length as the conventional slab AE device but with a much narrower width. For the numerical analysis, semiconductors in both designs are modeled with a length $L_0 = 800$ $\mu$m and a thickness $d = 50$ nm. First, we calculated the DC dissipated power at maximum gain in AE amplifiers, $P_{dc, Max} = \frac{qN\mu wd}{L_0}(\frac{v_a L_0}{\mu}+\frac{NqdL_0}{\epsilon_0+\epsilon_p})^2$, with an acoustic velocity $v_a = 6480$ m/s and a carrier concentration $N = 5\cdot10^{15}$ cm$^{-3}$ which is typical for heterogeneously integrated InGaAs. As shown in Fig. 6(b), $P_{dc, Max}$ significantly decreases as the semiconductor width narrows, which suggests superior potential for better thermal performances and power efficiencies. In section H of the Supplementary Material, we further show that even when accounting for additional propagation losses, AE waveguide amplifiers dissipate approximately 95\% less DC power than a slab AE amplifier for the same amount of gain. Second, we compare the nonlinearity in an AE slab 3-wave mixer and an AE waveguide 3-wave mixer construct with an InGaAs/AlScN/SiC heterostructure with the same amount of pump power. Here the nonlinearities are represented by the nonlinear AE coefficient $|\eta_3 u_1|$ of a three wave-mixing process for 3 GHz and 4 GHz Sezawa-like phonons\cite{Hackett2024GiantHeterostructures}(See Supplementary Material section I for calculation details). As shown in Fig. 6(c), the nonlinear AE coefficient $|\eta_3 u_1|$ is largely enhanced in waveguide mixers due to the higher phonon intensity per cross-sectional area. These comparing results highlight the AlScN/SiC phononic waveguides as an outstanding platform for AE devices. 

\begin{figure}[htbp]
\includegraphics[width=14cm]{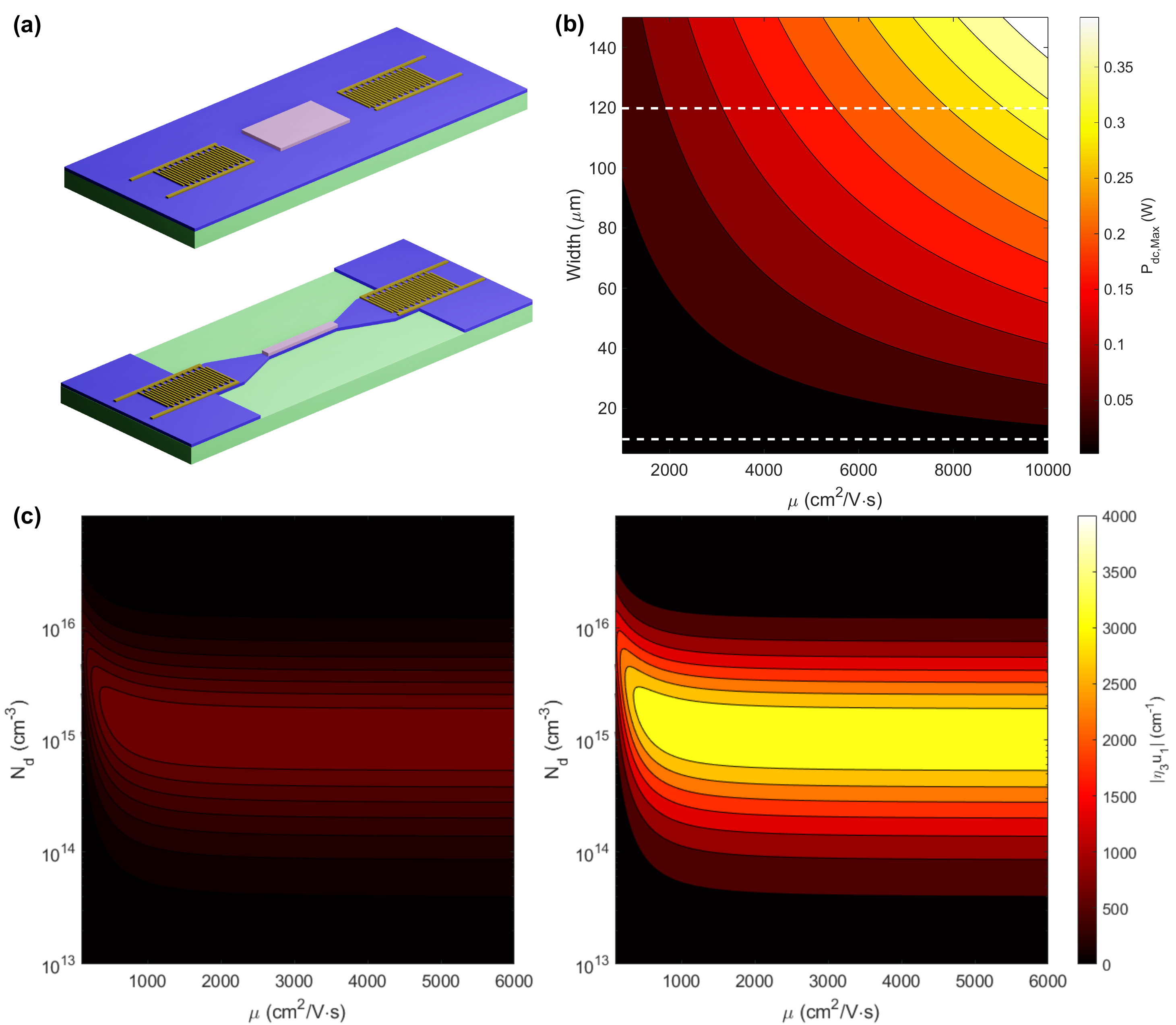}
\renewcommand{\figurename}{Fig.}
\caption{\label{fig:6}(a) Schematic illustrations of a conventional slab AE device (top) and a waveguide AE device (bottom). Pink blocks represent the semiconductor regions. (b) The dissipated DC power at maximum gain for a semiconductor width varied from 5 $\mu$m to 150 $\mu$m and mobility from 10 to 10000 $cm^{2}/V\cdot s$. White dashed lines represent AE amplifiers with estimated widths of the semiconductor in a slab and waveguide design illustrated in (a). (c) AE nonlinear coefficients $|\eta_3 u_1|$ of AE three-wave-mixers with a 120 $\mu$m wide slab (left) and a $4.8 \mu$m wide waveguide (right). }
\end{figure}

\section{Spin-phonon coupling}
Finally, we studied the spin-strain coupling between the divacancy centers in 4H-SiC and phonons in AlScN waveguides. Based on a previous study\cite{Whiteley2019SpinphononAcoustics}, phonon induced strain can couple with both $\Delta m_s = \pm1$ and $\Delta m_s = \pm2$ spin transitions of a divacancy center in SiC. In this section, we compare the coupling strength $\Omega_m$ induced by two phononic modes with a wavelength $\lambda = 1.6$ $\mu$m. The strain field $\epsilon_{ij}$ is obtained from COMSOL waveguide cross-sectional simulations and normalized by the number of phonons per wavelength $n = E/hf$, where $E$ is the phononic energy per wavelength. The strain coupling strength is correlated to the change of the zero-field splitting tensor $\Delta D$, which can be calculated from the product of the spin-strain coupling tensor $G$ and the strain tensor $\epsilon$. As a result, the spin-strain coupling strength $\Omega_m$ is proportional to $\frac{D_{xz}-iD_{xy}}{\sqrt2}$ for $\Delta m_s = \pm1$ spin transition and $\frac{1}{2}(D_{xx}-D_{yy})-iD_{xy} $ for $\Delta m_s = \pm2$ spin transition. The spatial distributions of these terms are plotted in Fig. 7. As shown in Fig 7 (a) and (b), the Sezawa-like mode displays a significantly stronger maximum coupling strength per phonon to the $\Delta m_s = \pm1$ spin transition of the $hh$ divacancy center in the SiC compared to the Rayleigh-like mode. In this case, both modes exhibit the position of the maximum coupling underneath the middle of the AlScN waveguide. In the other case illustrated in Fig. 7 (c) and (d), where these modes are coupled to the $\Delta m_s = \pm2$ spin transition, the position of the maximum coupling for the Sezawa-like mode is located at the side of the AlScN waveguide and maintains a much stronger coupling strength per phonon compared to the Rayleigh-like mode. These results suggest a great potential of our platform providing piezoelectric phonon coupled color centers embedded in SiC substrates. 

%For the $kk$ divacancy center, the Sezawa-like mode displays a stronger spin-strain coupling similarly (See Supplementary Materials Section X).

\begin{figure}
\includegraphics[width=12cm]{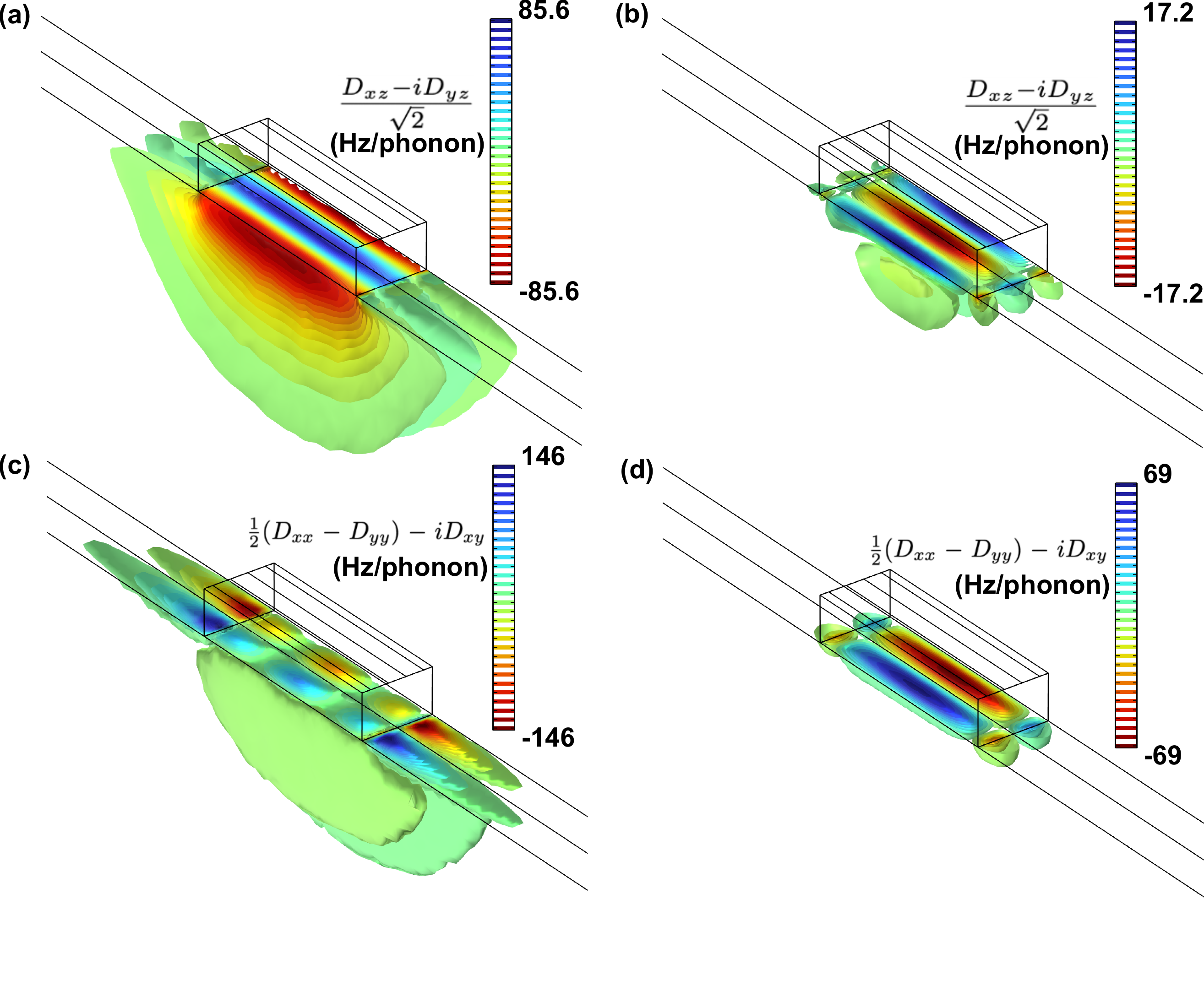}
\renewcommand{\figurename}{Fig.}
\caption{\label{fig:7} Spatial maps of the terms that are linearly related to the spin-strain coupling strengths of phonons with $hh$ divacancy centers in SiC. (a), (b) $\frac{D_{xz}-iD_{xy}}{\sqrt2}$ for (a) Sezawa-like and (b) Rayleigh-like modes. (c), (d) $\frac{1}{2}(D_{xx}-D_{yy})-iD_{xy} $ for (c) Sezawa-like and (d) Rayleigh-like modes. }
\end{figure}

\section{Conclusion}

In conclusion, we successfully designed, fabricated, and characterized phononic integrated waveguides on an AlScN/SiC platform, focusing on Sezawa-like and Rayleigh-like modes. Using impedance-matched interdigital transducers (IDTs), we efficiently excited these modes, with the Sezawa-like mode demonstrating a significantly higher electromechanical coupling coefficient. Slab and straight phononic waveguide delay lines were fabricated, tested and analyzed. Transmission measurements revealed higher propagation losses in the straight waveguides compared to the slab which is attributed to edge roughness from fabrication. Furthermore, we proposed AE waveguide amplifiers and wave mixers, theoretically evaluating the potential performance enhancements offered by the phononic waveguide design. The significantly reduced power dissipation and increased nonlinearity underscore the potential of AlScN/SiC phononic waveguides for high-efficiency AE devices. Finally, we showed that the Sezawa-like phonons in AlScN exhibits stronger spin-strain coupling to the divacancy centers in SiC compared to the Rayleigh-like phonons. These findings highlight the compelling potential of this platform and pave the way for broader applications of phononic devices in both classical and quantum domains.

\begin{acknowledgments}

This article is based on research sponsored in part by the Defense Advanced Research Projects Agency (DARPA) through a Young Faculty Award (YFA) under grant D23AP00174-00. The views and conclusions contained herein are those of the authors and should not be interpreted as necessarily representing the official policies or endorsements, either expressed or implied, by DARPA, the Department of the Interior, or the US Government. This work was performed, in part, at the Center for Integrated Nanotechnologies, an Office of Science User Facility operated for the U.S. Department of Energy, Office of Science. This work was also carried out, in part, at Sandia National Laboratories, a multimission laboratory managed and operated by National Technology and Engineering Solutions of Sandia, LLC, a wholly owned subsidiary of Honeywell International, Inc., for the U.S. Department of Energy’s National Nuclear Security Administration under Contract No. DE-NA-003525. This article is also based upon work supported by the U.S. Department of Energy, Office of Science, National Quantum Information Science Research Centers, Quantum Systems Accelerator. This paper describes objective technical results and analysis. Any subjective views or opinions that might be expressed in the paper do not necessarily represent the views of the U.S. Department of Energy or the U.S. Government. 
\end{acknowledgments}

%\nocite{*}

\bibliography{PhWG_0909.bib}% Produces the bibliography via BibTeX.

\begin{thebibliography}{59}%
\makeatletter
\providecommand \@ifxundefined [1]{%
 \@ifx{#1\undefined}
}%
\providecommand \@ifnum [1]{%
 \ifnum #1\expandafter \@firstoftwo
 \else \expandafter \@secondoftwo
 \fi
}%
\providecommand \@ifx [1]{%
 \ifx #1\expandafter \@firstoftwo
 \else \expandafter \@secondoftwo
 \fi
}%
\providecommand \natexlab [1]{#1}%
\providecommand \enquote  [1]{``#1''}%
\providecommand \bibnamefont  [1]{#1}%
\providecommand \bibfnamefont [1]{#1}%
\providecommand \citenamefont [1]{#1}%
\providecommand \href@noop [0]{\@secondoftwo}%
\providecommand \href [0]{\begingroup \@sanitize@url \@href}%
\providecommand \@href[1]{\@@startlink{#1}\@@href}%
\providecommand \@@href[1]{\endgroup#1\@@endlink}%
\providecommand \@sanitize@url [0]{\catcode `\\12\catcode `\$12\catcode `\&12\catcode `\#12\catcode `\^12\catcode `\_12\catcode `\%12\relax}%
\providecommand \@@startlink[1]{}%
\providecommand \@@endlink[0]{}%
\providecommand \url  [0]{\begingroup\@sanitize@url \@url }%
\providecommand \@url [1]{\endgroup\@href {#1}{\urlprefix }}%
\providecommand \urlprefix  [0]{URL }%
\providecommand \Eprint [0]{\href }%
\providecommand \doibase [0]{https://doi.org/}%
\providecommand \selectlanguage [0]{\@gobble}%
\providecommand \bibinfo  [0]{\@secondoftwo}%
\providecommand \bibfield  [0]{\@secondoftwo}%
\providecommand \translation [1]{[#1]}%
\providecommand \BibitemOpen [0]{}%
\providecommand \bibitemStop [0]{}%
\providecommand \bibitemNoStop [0]{.\EOS\space}%
\providecommand \EOS [0]{\spacefactor3000\relax}%
\providecommand \BibitemShut  [1]{\csname bibitem#1\endcsname}%
\let\auto@bib@innerbib\@empty
%</preamble>
\bibitem [{\citenamefont {{David Morgan}}\ and\ \citenamefont {{E.G.S. Paige}}(2007)}]{DavidMorgan2007SurfaceFilters}%
  \BibitemOpen
  \bibfield  {author} {\bibinfo {author} {\bibnamefont {{David Morgan}}}\ and\ \bibinfo {author} {\bibnamefont {{E.G.S. Paige}}},\ }\href {https://doi.org/10.1016/B978-0-12-372537-0.X5000-6} {\emph {\bibinfo {title} {{Surface Acoustic Wave Filters}}}}\ (\bibinfo  {publisher} {Elsevier},\ \bibinfo {year} {2007})\BibitemShut {NoStop}%
\bibitem [{\citenamefont {Hackett}\ \emph {et~al.}(2023{\natexlab{a}})\citenamefont {Hackett}, \citenamefont {Miller}, \citenamefont {Weatherred}, \citenamefont {Arterburn}, \citenamefont {Storey}, \citenamefont {Peake}, \citenamefont {Dominguez}, \citenamefont {Finnegan}, \citenamefont {Friedmann},\ and\ \citenamefont {Eichenfield}}]{Hackett2023Non-reciprocalOperation}%
  \BibitemOpen
  \bibfield  {author} {\bibinfo {author} {\bibfnamefont {L.}~\bibnamefont {Hackett}}, \bibinfo {author} {\bibfnamefont {M.}~\bibnamefont {Miller}}, \bibinfo {author} {\bibfnamefont {S.}~\bibnamefont {Weatherred}}, \bibinfo {author} {\bibfnamefont {S.}~\bibnamefont {Arterburn}}, \bibinfo {author} {\bibfnamefont {M.~J.}\ \bibnamefont {Storey}}, \bibinfo {author} {\bibfnamefont {G.}~\bibnamefont {Peake}}, \bibinfo {author} {\bibfnamefont {D.}~\bibnamefont {Dominguez}}, \bibinfo {author} {\bibfnamefont {P.~S.}\ \bibnamefont {Finnegan}}, \bibinfo {author} {\bibfnamefont {T.~A.}\ \bibnamefont {Friedmann}},\ and\ \bibinfo {author} {\bibfnamefont {M.}~\bibnamefont {Eichenfield}},\ }\bibfield  {title} {\bibinfo {title} {{Non-reciprocal acoustoelectric microwave amplifiers with net gain and low noise in continuous operation}},\ }\href {https://doi.org/10.1038/s41928-022-00908-6} {\bibfield  {journal} {\bibinfo  {journal} {Nature Electronics 2023 6:1}\ }\textbf {\bibinfo {volume} {6}},\ \bibinfo {pages} {76} (\bibinfo
  {year} {2023}{\natexlab{a}})}\BibitemShut {NoStop}%
\bibitem [{\citenamefont {Mansoorzare}\ and\ \citenamefont {Abdolvand}(2022{\natexlab{a}})}]{Mansoorzare2022MicromachinedProcessing}%
  \BibitemOpen
  \bibfield  {author} {\bibinfo {author} {\bibfnamefont {H.}~\bibnamefont {Mansoorzare}}\ and\ \bibinfo {author} {\bibfnamefont {R.}~\bibnamefont {Abdolvand}},\ }\bibfield  {title} {\bibinfo {title} {{Micromachined Heterostructured Lamb Mode Waveguides for Acoustoelectric Signal Processing}},\ }\href {https://doi.org/10.1109/TMTT.2022.3194723} {\bibfield  {journal} {\bibinfo  {journal} {IEEE Transactions on Microwave Theory and Techniques}\ }\textbf {\bibinfo {volume} {70}},\ \bibinfo {pages} {5195} (\bibinfo {year} {2022}{\natexlab{a}})}\BibitemShut {NoStop}%
\bibitem [{\citenamefont {Mansoorzare}\ and\ \citenamefont {Abdolvand}(2022{\natexlab{b}})}]{Mansoorzare2022Ultra-WidebandTransducers}%
  \BibitemOpen
  \bibfield  {author} {\bibinfo {author} {\bibfnamefont {H.}~\bibnamefont {Mansoorzare}}\ and\ \bibinfo {author} {\bibfnamefont {R.}~\bibnamefont {Abdolvand}},\ }\bibfield  {title} {\bibinfo {title} {{Ultra-Wideband Non-Reciprocal Micro-Acoustic Delay Lines with Slanted-Finger Interdigital Transducers}},\ }in\ \href {https://doi.org/10.1109/MEMS51670.2022.9699654} {\emph {\bibinfo {booktitle} {2022 IEEE 35th International Conference on Micro Electro Mechanical Systems Conference (MEMS)}}}\ (\bibinfo  {publisher} {IEEE},\ \bibinfo {year} {2022})\ pp.\ \bibinfo {pages} {1014--1017}\BibitemShut {NoStop}%
\bibitem [{\citenamefont {Hackett}\ \emph {et~al.}(2024{\natexlab{a}})\citenamefont {Hackett}, \citenamefont {Du}, \citenamefont {Miller}, \citenamefont {Smith}, \citenamefont {Santillan}, \citenamefont {Montoya}, \citenamefont {Reyna}, \citenamefont {Arterburn}, \citenamefont {Weatherred}, \citenamefont {Friedmann}, \citenamefont {Olsson},\ and\ \citenamefont {Eichenfield}}]{Hackett2024S-bandArchitecture}%
  \BibitemOpen
  \bibfield  {author} {\bibinfo {author} {\bibfnamefont {L.}~\bibnamefont {Hackett}}, \bibinfo {author} {\bibfnamefont {X.}~\bibnamefont {Du}}, \bibinfo {author} {\bibfnamefont {M.}~\bibnamefont {Miller}}, \bibinfo {author} {\bibfnamefont {B.}~\bibnamefont {Smith}}, \bibinfo {author} {\bibfnamefont {S.}~\bibnamefont {Santillan}}, \bibinfo {author} {\bibfnamefont {J.}~\bibnamefont {Montoya}}, \bibinfo {author} {\bibfnamefont {R.}~\bibnamefont {Reyna}}, \bibinfo {author} {\bibfnamefont {S.}~\bibnamefont {Arterburn}}, \bibinfo {author} {\bibfnamefont {S.}~\bibnamefont {Weatherred}}, \bibinfo {author} {\bibfnamefont {T.~A.}\ \bibnamefont {Friedmann}}, \bibinfo {author} {\bibfnamefont {R.~H.}\ \bibnamefont {Olsson}},\ and\ \bibinfo {author} {\bibfnamefont {M.}~\bibnamefont {Eichenfield}},\ }\bibfield  {title} {\bibinfo {title} {{S-band acoustoelectric amplifier in an InGaAs-AlScN-SiC architecture}},\ }\bibfield  {journal} {\bibinfo  {journal} {Applied Physics Letters}\ }\textbf {\bibinfo {volume} {124}},\ \href
  {https://doi.org/10.1063/5.0178912} {10.1063/5.0178912} (\bibinfo {year} {2024}{\natexlab{a}})\BibitemShut {NoStop}%
\bibitem [{\citenamefont {Hackett}\ \emph {et~al.}(2024{\natexlab{b}})\citenamefont {Hackett}, \citenamefont {Koppa}, \citenamefont {Smith}, \citenamefont {Miller}, \citenamefont {Santillan}, \citenamefont {Weatherred}, \citenamefont {Arterburn}, \citenamefont {Friedmann}, \citenamefont {Otterstrom},\ and\ \citenamefont {Eichenfield}}]{Hackett2024GiantHeterostructures}%
  \BibitemOpen
  \bibfield  {author} {\bibinfo {author} {\bibfnamefont {L.}~\bibnamefont {Hackett}}, \bibinfo {author} {\bibfnamefont {M.}~\bibnamefont {Koppa}}, \bibinfo {author} {\bibfnamefont {B.}~\bibnamefont {Smith}}, \bibinfo {author} {\bibfnamefont {M.}~\bibnamefont {Miller}}, \bibinfo {author} {\bibfnamefont {S.}~\bibnamefont {Santillan}}, \bibinfo {author} {\bibfnamefont {S.}~\bibnamefont {Weatherred}}, \bibinfo {author} {\bibfnamefont {S.}~\bibnamefont {Arterburn}}, \bibinfo {author} {\bibfnamefont {T.~A.}\ \bibnamefont {Friedmann}}, \bibinfo {author} {\bibfnamefont {N.}~\bibnamefont {Otterstrom}},\ and\ \bibinfo {author} {\bibfnamefont {M.}~\bibnamefont {Eichenfield}},\ }\bibfield  {title} {\bibinfo {title} {{Giant electron-mediated phononic nonlinearity in semiconductor–piezoelectric heterostructures}},\ }\href {https://doi.org/10.1038/s41563-024-01882-4} {\bibfield  {journal} {\bibinfo  {journal} {Nature Materials 2024}\ ,\ \bibinfo {pages} {1}} (\bibinfo {year} {2024}{\natexlab{b}})}\BibitemShut {NoStop}%
\bibitem [{\citenamefont {Mansoorzare}\ and\ \citenamefont {Abdolvand}(2023)}]{Mansoorzare2023Acoustoelectric-DrivenWaveguides}%
  \BibitemOpen
  \bibfield  {author} {\bibinfo {author} {\bibfnamefont {H.}~\bibnamefont {Mansoorzare}}\ and\ \bibinfo {author} {\bibfnamefont {R.}~\bibnamefont {Abdolvand}},\ }\bibfield  {title} {\bibinfo {title} {{Acoustoelectric-Driven Frequency Mixing in Micromachined Lithium Niobate on Silicon Waveguides}},\ }in\ \href {https://doi.org/10.1109/MEMS49605.2023.10052257} {\emph {\bibinfo {booktitle} {2023 IEEE 36th International Conference on Micro Electro Mechanical Systems (MEMS)}}}\ (\bibinfo  {publisher} {IEEE},\ \bibinfo {year} {2023})\ pp.\ \bibinfo {pages} {1183--1185}\BibitemShut {NoStop}%
\bibitem [{\citenamefont {Falk}\ \emph {et~al.}(2014)\citenamefont {Falk}, \citenamefont {Klimov}, \citenamefont {Buckley}, \citenamefont {Iv{\'{a}}dy}, \citenamefont {Abrikosov}, \citenamefont {Calusine}, \citenamefont {Koehl}, \citenamefont {Gali},\ and\ \citenamefont {Awschalom}}]{Falk2014ElectricallyCenters}%
  \BibitemOpen
  \bibfield  {author} {\bibinfo {author} {\bibfnamefont {A.~L.}\ \bibnamefont {Falk}}, \bibinfo {author} {\bibfnamefont {P.~V.}\ \bibnamefont {Klimov}}, \bibinfo {author} {\bibfnamefont {B.~B.}\ \bibnamefont {Buckley}}, \bibinfo {author} {\bibfnamefont {V.}~\bibnamefont {Iv{\'{a}}dy}}, \bibinfo {author} {\bibfnamefont {I.~A.}\ \bibnamefont {Abrikosov}}, \bibinfo {author} {\bibfnamefont {G.}~\bibnamefont {Calusine}}, \bibinfo {author} {\bibfnamefont {W.~F.}\ \bibnamefont {Koehl}}, \bibinfo {author} {\bibfnamefont {Ã.}~\bibnamefont {Gali}},\ and\ \bibinfo {author} {\bibfnamefont {D.~D.}\ \bibnamefont {Awschalom}},\ }\bibfield  {title} {\bibinfo {title} {{Electrically and mechanically tunable electron spins in silicon carbide color centers}},\ }\href {https://doi.org/10.1103/PHYSREVLETT.112.187601/FIGURES/4/MEDIUM} {\bibfield  {journal} {\bibinfo  {journal} {Physical Review Letters}\ }\textbf {\bibinfo {volume} {112}},\ \bibinfo {pages} {187601} (\bibinfo {year} {2014})}\BibitemShut {NoStop}%
\bibitem [{\citenamefont {Udvarhelyi}\ and\ \citenamefont {Gali}(2018)}]{Udvarhelyi2018iAbCarbide}%
  \BibitemOpen
  \bibfield  {author} {\bibinfo {author} {\bibfnamefont {P.}~\bibnamefont {Udvarhelyi}}\ and\ \bibinfo {author} {\bibfnamefont {A.}~\bibnamefont {Gali}},\ }\bibfield  {title} {\bibinfo {title} {{<i>Ab Initio</i> Spin-Strain Coupling Parameters of Divacancy Qubits in Silicon Carbide}},\ }\href {https://doi.org/10.1103/PhysRevApplied.10.054010} {\bibfield  {journal} {\bibinfo  {journal} {Physical Review Applied}\ }\textbf {\bibinfo {volume} {10}},\ \bibinfo {pages} {054010} (\bibinfo {year} {2018})}\BibitemShut {NoStop}%
\bibitem [{\citenamefont {Whiteley}\ \emph {et~al.}(2019)\citenamefont {Whiteley}, \citenamefont {Wolfowicz}, \citenamefont {Anderson}, \citenamefont {Bourassa}, \citenamefont {Ma}, \citenamefont {Ye}, \citenamefont {Koolstra}, \citenamefont {Satzinger}, \citenamefont {Holt}, \citenamefont {Heremans}, \citenamefont {Cleland}, \citenamefont {Schuster}, \citenamefont {Galli},\ and\ \citenamefont {Awschalom}}]{Whiteley2019SpinphononAcoustics}%
  \BibitemOpen
  \bibfield  {author} {\bibinfo {author} {\bibfnamefont {S.~J.}\ \bibnamefont {Whiteley}}, \bibinfo {author} {\bibfnamefont {G.}~\bibnamefont {Wolfowicz}}, \bibinfo {author} {\bibfnamefont {C.~P.}\ \bibnamefont {Anderson}}, \bibinfo {author} {\bibfnamefont {A.}~\bibnamefont {Bourassa}}, \bibinfo {author} {\bibfnamefont {H.}~\bibnamefont {Ma}}, \bibinfo {author} {\bibfnamefont {M.}~\bibnamefont {Ye}}, \bibinfo {author} {\bibfnamefont {G.}~\bibnamefont {Koolstra}}, \bibinfo {author} {\bibfnamefont {K.~J.}\ \bibnamefont {Satzinger}}, \bibinfo {author} {\bibfnamefont {M.~V.}\ \bibnamefont {Holt}}, \bibinfo {author} {\bibfnamefont {F.~J.}\ \bibnamefont {Heremans}}, \bibinfo {author} {\bibfnamefont {A.~N.}\ \bibnamefont {Cleland}}, \bibinfo {author} {\bibfnamefont {D.~I.}\ \bibnamefont {Schuster}}, \bibinfo {author} {\bibfnamefont {G.}~\bibnamefont {Galli}},\ and\ \bibinfo {author} {\bibfnamefont {D.~D.}\ \bibnamefont {Awschalom}},\ }\bibfield  {title} {\bibinfo {title} {{Spin–phonon interactions in silicon
  carbide addressed by Gaussian acoustics}},\ }\href {https://doi.org/10.1038/s41567-019-0420-0} {\bibfield  {journal} {\bibinfo  {journal} {Nature Physics}\ }\textbf {\bibinfo {volume} {15}},\ \bibinfo {pages} {490} (\bibinfo {year} {2019})}\BibitemShut {NoStop}%
\bibitem [{\citenamefont {Maity}\ \emph {et~al.}(2020)\citenamefont {Maity}, \citenamefont {Shao}, \citenamefont {Bogdanovi{\'{c}}}, \citenamefont {Meesala}, \citenamefont {Sohn}, \citenamefont {Sinclair}, \citenamefont {Pingault}, \citenamefont {Chalupnik}, \citenamefont {Chia}, \citenamefont {Zheng}, \citenamefont {Lai},\ and\ \citenamefont {Lon{\v{c}}ar}}]{Maity2020CoherentDiamond}%
  \BibitemOpen
  \bibfield  {author} {\bibinfo {author} {\bibfnamefont {S.}~\bibnamefont {Maity}}, \bibinfo {author} {\bibfnamefont {L.}~\bibnamefont {Shao}}, \bibinfo {author} {\bibfnamefont {S.}~\bibnamefont {Bogdanovi{\'{c}}}}, \bibinfo {author} {\bibfnamefont {S.}~\bibnamefont {Meesala}}, \bibinfo {author} {\bibfnamefont {Y.-I.}\ \bibnamefont {Sohn}}, \bibinfo {author} {\bibfnamefont {N.}~\bibnamefont {Sinclair}}, \bibinfo {author} {\bibfnamefont {B.}~\bibnamefont {Pingault}}, \bibinfo {author} {\bibfnamefont {M.}~\bibnamefont {Chalupnik}}, \bibinfo {author} {\bibfnamefont {C.}~\bibnamefont {Chia}}, \bibinfo {author} {\bibfnamefont {L.}~\bibnamefont {Zheng}}, \bibinfo {author} {\bibfnamefont {K.}~\bibnamefont {Lai}},\ and\ \bibinfo {author} {\bibfnamefont {M.}~\bibnamefont {Lon{\v{c}}ar}},\ }\bibfield  {title} {\bibinfo {title} {{Coherent acoustic control of a single silicon vacancy spin in diamond}},\ }\href {https://doi.org/10.1038/s41467-019-13822-x} {\bibfield  {journal} {\bibinfo  {journal} {Nature Communications}\
  }\textbf {\bibinfo {volume} {11}},\ \bibinfo {pages} {193} (\bibinfo {year} {2020})}\BibitemShut {NoStop}%
\bibitem [{\citenamefont {Dietz}\ \emph {et~al.}(2023)\citenamefont {Dietz}, \citenamefont {Jiang}, \citenamefont {Day}, \citenamefont {Bhave},\ and\ \citenamefont {Hu}}]{Dietz2023Spin-acousticCarbide}%
  \BibitemOpen
  \bibfield  {author} {\bibinfo {author} {\bibfnamefont {J.~R.}\ \bibnamefont {Dietz}}, \bibinfo {author} {\bibfnamefont {B.}~\bibnamefont {Jiang}}, \bibinfo {author} {\bibfnamefont {A.~M.}\ \bibnamefont {Day}}, \bibinfo {author} {\bibfnamefont {S.~A.}\ \bibnamefont {Bhave}},\ and\ \bibinfo {author} {\bibfnamefont {E.~L.}\ \bibnamefont {Hu}},\ }\bibfield  {title} {\bibinfo {title} {{Spin-acoustic control of silicon vacancies in 4H silicon carbide}},\ }\href {https://doi.org/10.1038/s41928-023-01029-4} {\bibfield  {journal} {\bibinfo  {journal} {Nature Electronics}\ }\textbf {\bibinfo {volume} {6}},\ \bibinfo {pages} {739} (\bibinfo {year} {2023})}\BibitemShut {NoStop}%
\bibitem [{\citenamefont {Kuruma}\ \emph {et~al.}(2024)\citenamefont {Kuruma}, \citenamefont {Pingault}, \citenamefont {Chia}, \citenamefont {Haas}, \citenamefont {Joe}, \citenamefont {Assumpcao}, \citenamefont {Ding}, \citenamefont {Jin}, \citenamefont {Xin}, \citenamefont {Yeh}, \citenamefont {Sinclair},\ and\ \citenamefont {Lon{\v{c}}ar}}]{Kuruma2024ControllingCrystals}%
  \BibitemOpen
  \bibfield  {author} {\bibinfo {author} {\bibfnamefont {K.}~\bibnamefont {Kuruma}}, \bibinfo {author} {\bibfnamefont {B.}~\bibnamefont {Pingault}}, \bibinfo {author} {\bibfnamefont {C.}~\bibnamefont {Chia}}, \bibinfo {author} {\bibfnamefont {M.}~\bibnamefont {Haas}}, \bibinfo {author} {\bibfnamefont {G.~D.}\ \bibnamefont {Joe}}, \bibinfo {author} {\bibfnamefont {D.~R.}\ \bibnamefont {Assumpcao}}, \bibinfo {author} {\bibfnamefont {S.~W.}\ \bibnamefont {Ding}}, \bibinfo {author} {\bibfnamefont {C.}~\bibnamefont {Jin}}, \bibinfo {author} {\bibfnamefont {C.~J.}\ \bibnamefont {Xin}}, \bibinfo {author} {\bibfnamefont {M.}~\bibnamefont {Yeh}}, \bibinfo {author} {\bibfnamefont {N.}~\bibnamefont {Sinclair}},\ and\ \bibinfo {author} {\bibfnamefont {M.}~\bibnamefont {Lon{\v{c}}ar}},\ }\bibfield  {title} {\bibinfo {title} {{Controlling interactions between high-frequency phonons and single quantum systems using phononic crystals}},\ }\href {https://doi.org/10.1038/s41567-024-02697-5} {\bibfield  {journal} {\bibinfo
  {journal} {Nature Physics 2024}\ ,\ \bibinfo {pages} {1}} (\bibinfo {year} {2024})}\BibitemShut {NoStop}%
\bibitem [{\citenamefont {Clark}\ \emph {et~al.}(2024)\citenamefont {Clark}, \citenamefont {Raniwala}, \citenamefont {Koppa}, \citenamefont {Chen}, \citenamefont {Leenheer}, \citenamefont {Zimmermann}, \citenamefont {Dong}, \citenamefont {Li}, \citenamefont {Wen}, \citenamefont {Dominguez}, \citenamefont {Trusheim}, \citenamefont {Gilbert}, \citenamefont {Eichenfield},\ and\ \citenamefont {Englund}}]{Clark2024NanoelectromechanicalChip}%
  \BibitemOpen
  \bibfield  {author} {\bibinfo {author} {\bibfnamefont {G.}~\bibnamefont {Clark}}, \bibinfo {author} {\bibfnamefont {H.}~\bibnamefont {Raniwala}}, \bibinfo {author} {\bibfnamefont {M.}~\bibnamefont {Koppa}}, \bibinfo {author} {\bibfnamefont {K.}~\bibnamefont {Chen}}, \bibinfo {author} {\bibfnamefont {A.}~\bibnamefont {Leenheer}}, \bibinfo {author} {\bibfnamefont {M.}~\bibnamefont {Zimmermann}}, \bibinfo {author} {\bibfnamefont {M.}~\bibnamefont {Dong}}, \bibinfo {author} {\bibfnamefont {L.}~\bibnamefont {Li}}, \bibinfo {author} {\bibfnamefont {Y.~H.}\ \bibnamefont {Wen}}, \bibinfo {author} {\bibfnamefont {D.}~\bibnamefont {Dominguez}}, \bibinfo {author} {\bibfnamefont {M.}~\bibnamefont {Trusheim}}, \bibinfo {author} {\bibfnamefont {G.}~\bibnamefont {Gilbert}}, \bibinfo {author} {\bibfnamefont {M.}~\bibnamefont {Eichenfield}},\ and\ \bibinfo {author} {\bibfnamefont {D.}~\bibnamefont {Englund}},\ }\bibfield  {title} {\bibinfo {title} {{Nanoelectromechanical Control of Spin-Photon Interfaces in a Hybrid Quantum
  System on Chip}},\ }\href {https://doi.org/10.1021/ACS.NANOLETT.3C04301/ASSET/IMAGES/LARGE/NL3C04301{\_}0005.JPEG} {\bibfield  {journal} {\bibinfo  {journal} {Nano Letters}\ }\textbf {\bibinfo {volume} {24}},\ \bibinfo {pages} {1316} (\bibinfo {year} {2024})}\BibitemShut {NoStop}%
\bibitem [{\citenamefont {Qiao}\ \emph {et~al.}(2023)\citenamefont {Qiao}, \citenamefont {{Dumur}}, \citenamefont {Andersson}, \citenamefont {Yan}, \citenamefont {Chou}, \citenamefont {Grebel}, \citenamefont {Conner}, \citenamefont {Joshi}, \citenamefont {Miller}, \citenamefont {Povey}, \citenamefont {Wu},\ and\ \citenamefont {Cleland}}]{Qiao2023SplittingComputing}%
  \BibitemOpen
  \bibfield  {author} {\bibinfo {author} {\bibfnamefont {H.}~\bibnamefont {Qiao}}, \bibinfo {author} {\bibnamefont {{Dumur}}}, \bibinfo {author} {\bibfnamefont {G.}~\bibnamefont {Andersson}}, \bibinfo {author} {\bibfnamefont {H.}~\bibnamefont {Yan}}, \bibinfo {author} {\bibfnamefont {M.~H.}\ \bibnamefont {Chou}}, \bibinfo {author} {\bibfnamefont {J.}~\bibnamefont {Grebel}}, \bibinfo {author} {\bibfnamefont {C.~R.}\ \bibnamefont {Conner}}, \bibinfo {author} {\bibfnamefont {Y.~J.}\ \bibnamefont {Joshi}}, \bibinfo {author} {\bibfnamefont {J.~M.}\ \bibnamefont {Miller}}, \bibinfo {author} {\bibfnamefont {R.~G.}\ \bibnamefont {Povey}}, \bibinfo {author} {\bibfnamefont {X.}~\bibnamefont {Wu}},\ and\ \bibinfo {author} {\bibfnamefont {A.~N.}\ \bibnamefont {Cleland}},\ }\bibfield  {title} {\bibinfo {title} {{Splitting phonons: Building a platform for linear mechanical quantum computing}},\ }\href {https://doi.org/10.1126/SCIENCE.ADG8715/SUPPL{\_}FILE/SCIENCE.ADG8715{\_}SM.PDF} {\bibfield  {journal} {\bibinfo
  {journal} {Science}\ }\textbf {\bibinfo {volume} {380}},\ \bibinfo {pages} {1030} (\bibinfo {year} {2023})}\BibitemShut {NoStop}%
\bibitem [{\citenamefont {Boisen}\ \emph {et~al.}(2018)\citenamefont {Boisen}, \citenamefont {Dohn}, \citenamefont {Keller}, \citenamefont {Pechal}, \citenamefont {Arrangoiz-Arriola},\ and\ \citenamefont {Safavi-Naeini}}]{Boisen2018SuperconductingStorage}%
  \BibitemOpen
  \bibfield  {author} {\bibinfo {author} {\bibfnamefont {A.}~\bibnamefont {Boisen}}, \bibinfo {author} {\bibfnamefont {S.}~\bibnamefont {Dohn}}, \bibinfo {author} {\bibfnamefont {S.~S.}\ \bibnamefont {Keller}}, \bibinfo {author} {\bibfnamefont {M.}~\bibnamefont {Pechal}}, \bibinfo {author} {\bibfnamefont {P.}~\bibnamefont {Arrangoiz-Arriola}},\ and\ \bibinfo {author} {\bibfnamefont {A.~H.}\ \bibnamefont {Safavi-Naeini}},\ }\bibfield  {title} {\bibinfo {title} {{Superconducting circuit quantum computing with nanomechanical resonators as storage}},\ }\href {https://doi.org/10.1088/2058-9565/AADC6C} {\bibfield  {journal} {\bibinfo  {journal} {Quantum Science and Technology}\ }\textbf {\bibinfo {volume} {4}},\ \bibinfo {pages} {015006} (\bibinfo {year} {2018})}\BibitemShut {NoStop}%
\bibitem [{\citenamefont {Chu}\ \emph {et~al.}(2018)\citenamefont {Chu}, \citenamefont {Kharel}, \citenamefont {Yoon}, \citenamefont {Frunzio}, \citenamefont {Rakich},\ and\ \citenamefont {Schoelkopf}}]{Chu2018CreationResonator}%
  \BibitemOpen
  \bibfield  {author} {\bibinfo {author} {\bibfnamefont {Y.}~\bibnamefont {Chu}}, \bibinfo {author} {\bibfnamefont {P.}~\bibnamefont {Kharel}}, \bibinfo {author} {\bibfnamefont {T.}~\bibnamefont {Yoon}}, \bibinfo {author} {\bibfnamefont {L.}~\bibnamefont {Frunzio}}, \bibinfo {author} {\bibfnamefont {P.~T.}\ \bibnamefont {Rakich}},\ and\ \bibinfo {author} {\bibfnamefont {R.~J.}\ \bibnamefont {Schoelkopf}},\ }\bibfield  {title} {\bibinfo {title} {{Creation and control of multi-phonon Fock states in a bulk acoustic-wave resonator}},\ }\href {https://doi.org/10.1038/s41586-018-0717-7} {\bibfield  {journal} {\bibinfo  {journal} {Nature 2018 563:7733}\ }\textbf {\bibinfo {volume} {563}},\ \bibinfo {pages} {666} (\bibinfo {year} {2018})}\BibitemShut {NoStop}%
\bibitem [{\citenamefont {Hann}\ \emph {et~al.}(2019)\citenamefont {Hann}, \citenamefont {Zou}, \citenamefont {Zhang}, \citenamefont {Chu}, \citenamefont {Schoelkopf}, \citenamefont {Girvin},\ and\ \citenamefont {Jiang}}]{Hann2019Hardware-EfficientSystems}%
  \BibitemOpen
  \bibfield  {author} {\bibinfo {author} {\bibfnamefont {C.~T.}\ \bibnamefont {Hann}}, \bibinfo {author} {\bibfnamefont {C.~L.}\ \bibnamefont {Zou}}, \bibinfo {author} {\bibfnamefont {Y.}~\bibnamefont {Zhang}}, \bibinfo {author} {\bibfnamefont {Y.}~\bibnamefont {Chu}}, \bibinfo {author} {\bibfnamefont {R.~J.}\ \bibnamefont {Schoelkopf}}, \bibinfo {author} {\bibfnamefont {S.~M.}\ \bibnamefont {Girvin}},\ and\ \bibinfo {author} {\bibfnamefont {L.}~\bibnamefont {Jiang}},\ }\bibfield  {title} {\bibinfo {title} {{Hardware-Efficient Quantum Random Access Memory with Hybrid Quantum Acoustic Systems}},\ }\href {https://doi.org/10.1103/PHYSREVLETT.123.250501/FIGURES/4/MEDIUM} {\bibfield  {journal} {\bibinfo  {journal} {Physical Review Letters}\ }\textbf {\bibinfo {volume} {123}},\ \bibinfo {pages} {250501} (\bibinfo {year} {2019})}\BibitemShut {NoStop}%
\bibitem [{\citenamefont {Yang}\ \emph {et~al.}(2025)\citenamefont {Yang}, \citenamefont {Xie},\ and\ \citenamefont {Tang}}]{Yang2025FluorescenceMicrocavity}%
  \BibitemOpen
  \bibfield  {author} {\bibinfo {author} {\bibfnamefont {L.}~\bibnamefont {Yang}}, \bibinfo {author} {\bibfnamefont {J.}~\bibnamefont {Xie}},\ and\ \bibinfo {author} {\bibfnamefont {H.~X.}\ \bibnamefont {Tang}},\ }\bibfield  {title} {\bibinfo {title} {{Fluorescence enabled phonon counting in an erbium-doped piezo-optomechanical microcavity}},\ }\bibfield  {journal} {\bibinfo  {journal} {Nanophotonics}\ }\href {https://doi.org/10.1515/NANOPH-2024-0400} {10.1515/NANOPH-2024-0400} (\bibinfo {year} {2025})\BibitemShut {NoStop}%
\bibitem [{\citenamefont {Dong}\ \emph {et~al.}(2012)\citenamefont {Dong}, \citenamefont {Cui}, \citenamefont {Zhang}, \citenamefont {Yan}, \citenamefont {Cheng}, \citenamefont {Lu}, \citenamefont {Habraken}, \citenamefont {Stannigel}, \citenamefont {Lukin}, \citenamefont {Zoller},\ and\ \citenamefont {Rabl}}]{Dong2012ContinuousNetworks}%
  \BibitemOpen
  \bibfield  {author} {\bibinfo {author} {\bibfnamefont {Z.}~\bibnamefont {Dong}}, \bibinfo {author} {\bibfnamefont {L.}~\bibnamefont {Cui}}, \bibinfo {author} {\bibfnamefont {G.}~\bibnamefont {Zhang}}, \bibinfo {author} {\bibfnamefont {G.-A.}\ \bibnamefont {Yan}}, \bibinfo {author} {\bibfnamefont {W.-Q.}\ \bibnamefont {Cheng}}, \bibinfo {author} {\bibfnamefont {H.}~\bibnamefont {Lu}}, \bibinfo {author} {\bibfnamefont {S.~J.~M.}\ \bibnamefont {Habraken}}, \bibinfo {author} {\bibfnamefont {K.}~\bibnamefont {Stannigel}}, \bibinfo {author} {\bibfnamefont {M.~D.}\ \bibnamefont {Lukin}}, \bibinfo {author} {\bibfnamefont {P.}~\bibnamefont {Zoller}},\ and\ \bibinfo {author} {\bibfnamefont {P.}~\bibnamefont {Rabl}},\ }\bibfield  {title} {\bibinfo {title} {{Continuous mode cooling and phonon routers for phononic quantum networks}},\ }\href {https://doi.org/10.1088/1367-2630/14/11/115004} {\bibfield  {journal} {\bibinfo  {journal} {New Journal of Physics}\ }\textbf {\bibinfo {volume} {14}},\ \bibinfo {pages} {115004}
  (\bibinfo {year} {2012})}\BibitemShut {NoStop}%
\bibitem [{\citenamefont {Golter}\ \emph {et~al.}(2016)\citenamefont {Golter}, \citenamefont {Oo}, \citenamefont {Amezcua}, \citenamefont {Lekavicius}, \citenamefont {Stewart},\ and\ \citenamefont {Wang}}]{Golter2016CouplingState}%
  \BibitemOpen
  \bibfield  {author} {\bibinfo {author} {\bibfnamefont {D.~A.}\ \bibnamefont {Golter}}, \bibinfo {author} {\bibfnamefont {T.}~\bibnamefont {Oo}}, \bibinfo {author} {\bibfnamefont {M.}~\bibnamefont {Amezcua}}, \bibinfo {author} {\bibfnamefont {I.}~\bibnamefont {Lekavicius}}, \bibinfo {author} {\bibfnamefont {K.~A.}\ \bibnamefont {Stewart}},\ and\ \bibinfo {author} {\bibfnamefont {H.}~\bibnamefont {Wang}},\ }\bibfield  {title} {\bibinfo {title} {{Coupling a surface acoustic wave to an electron spin in diamond via a dark state}},\ }\href {https://doi.org/10.1103/PHYSREVX.6.041060/FIGURES/6/MEDIUM} {\bibfield  {journal} {\bibinfo  {journal} {Physical Review X}\ }\textbf {\bibinfo {volume} {6}},\ \bibinfo {pages} {041060} (\bibinfo {year} {2016})}\BibitemShut {NoStop}%
\bibitem [{\citenamefont {Chu}\ \emph {et~al.}(2017)\citenamefont {Chu}, \citenamefont {Kharel}, \citenamefont {Renninger}, \citenamefont {Burkhart}, \citenamefont {Frunzio}, \citenamefont {Rakich},\ and\ \citenamefont {Schoelkopf}}]{Chu2017QuantumQubits}%
  \BibitemOpen
  \bibfield  {author} {\bibinfo {author} {\bibfnamefont {Y.}~\bibnamefont {Chu}}, \bibinfo {author} {\bibfnamefont {P.}~\bibnamefont {Kharel}}, \bibinfo {author} {\bibfnamefont {W.~H.}\ \bibnamefont {Renninger}}, \bibinfo {author} {\bibfnamefont {L.~D.}\ \bibnamefont {Burkhart}}, \bibinfo {author} {\bibfnamefont {L.}~\bibnamefont {Frunzio}}, \bibinfo {author} {\bibfnamefont {P.~T.}\ \bibnamefont {Rakich}},\ and\ \bibinfo {author} {\bibfnamefont {R.~J.}\ \bibnamefont {Schoelkopf}},\ }\bibfield  {title} {\bibinfo {title} {{Quantum acoustics with superconducting qubits}},\ }\href {https://doi.org/10.1126/SCIENCE.AAO1511/SUPPL{\_}FILE/AAO1511{\_}CHU{\_}SM.PDF} {\bibfield  {journal} {\bibinfo  {journal} {Science}\ }\textbf {\bibinfo {volume} {358}},\ \bibinfo {pages} {199} (\bibinfo {year} {2017})}\BibitemShut {NoStop}%
\bibitem [{\citenamefont {Manenti}\ \emph {et~al.}(2017)\citenamefont {Manenti}, \citenamefont {Kockum}, \citenamefont {Patterson}, \citenamefont {Behrle}, \citenamefont {Rahamim}, \citenamefont {Tancredi}, \citenamefont {Nori},\ and\ \citenamefont {Leek}}]{Manenti2017CircuitWaves}%
  \BibitemOpen
  \bibfield  {author} {\bibinfo {author} {\bibfnamefont {R.}~\bibnamefont {Manenti}}, \bibinfo {author} {\bibfnamefont {A.~F.}\ \bibnamefont {Kockum}}, \bibinfo {author} {\bibfnamefont {A.}~\bibnamefont {Patterson}}, \bibinfo {author} {\bibfnamefont {T.}~\bibnamefont {Behrle}}, \bibinfo {author} {\bibfnamefont {J.}~\bibnamefont {Rahamim}}, \bibinfo {author} {\bibfnamefont {G.}~\bibnamefont {Tancredi}}, \bibinfo {author} {\bibfnamefont {F.}~\bibnamefont {Nori}},\ and\ \bibinfo {author} {\bibfnamefont {P.~J.}\ \bibnamefont {Leek}},\ }\bibfield  {title} {\bibinfo {title} {{Circuit quantum acoustodynamics with surface acoustic waves}},\ }\href {https://doi.org/10.1038/s41467-017-01063-9} {\bibfield  {journal} {\bibinfo  {journal} {Nature Communications}\ }\textbf {\bibinfo {volume} {8}},\ \bibinfo {pages} {975} (\bibinfo {year} {2017})}\BibitemShut {NoStop}%
\bibitem [{\citenamefont {Arrangoiz-Arriola}\ \emph {et~al.}(2018)\citenamefont {Arrangoiz-Arriola}, \citenamefont {Wollack}, \citenamefont {Pechal}, \citenamefont {Witmer}, \citenamefont {Hill},\ and\ \citenamefont {Safavi-Naeini}}]{Arrangoiz-Arriola2018CouplingCavity}%
  \BibitemOpen
  \bibfield  {author} {\bibinfo {author} {\bibfnamefont {P.}~\bibnamefont {Arrangoiz-Arriola}}, \bibinfo {author} {\bibfnamefont {E.~A.}\ \bibnamefont {Wollack}}, \bibinfo {author} {\bibfnamefont {M.}~\bibnamefont {Pechal}}, \bibinfo {author} {\bibfnamefont {J.~D.}\ \bibnamefont {Witmer}}, \bibinfo {author} {\bibfnamefont {J.~T.}\ \bibnamefont {Hill}},\ and\ \bibinfo {author} {\bibfnamefont {A.~H.}\ \bibnamefont {Safavi-Naeini}},\ }\bibfield  {title} {\bibinfo {title} {{Coupling a Superconducting Quantum Circuit to a Phononic Crystal Defect Cavity}},\ }\href {https://doi.org/10.1103/PHYSREVX.8.031007/FIGURES/8/MEDIUM} {\bibfield  {journal} {\bibinfo  {journal} {Physical Review X}\ }\textbf {\bibinfo {volume} {8}},\ \bibinfo {pages} {031007} (\bibinfo {year} {2018})}\BibitemShut {NoStop}%
\bibitem [{\citenamefont {Moores}\ \emph {et~al.}(2018)\citenamefont {Moores}, \citenamefont {Sletten}, \citenamefont {Viennot},\ and\ \citenamefont {Lehnert}}]{Moores2018CavityRegime}%
  \BibitemOpen
  \bibfield  {author} {\bibinfo {author} {\bibfnamefont {B.~A.}\ \bibnamefont {Moores}}, \bibinfo {author} {\bibfnamefont {L.~R.}\ \bibnamefont {Sletten}}, \bibinfo {author} {\bibfnamefont {J.~J.}\ \bibnamefont {Viennot}},\ and\ \bibinfo {author} {\bibfnamefont {K.~W.}\ \bibnamefont {Lehnert}},\ }\bibfield  {title} {\bibinfo {title} {{Cavity Quantum Acoustic Device in the Multimode Strong Coupling Regime}},\ }\href {https://doi.org/10.1103/PHYSREVLETT.120.227701/FIGURES/4/MEDIUM} {\bibfield  {journal} {\bibinfo  {journal} {Physical Review Letters}\ }\textbf {\bibinfo {volume} {120}},\ \bibinfo {pages} {227701} (\bibinfo {year} {2018})}\BibitemShut {NoStop}%
\bibitem [{\citenamefont {Lemonde}\ \emph {et~al.}(2018)\citenamefont {Lemonde}, \citenamefont {Meesala}, \citenamefont {Sipahigil}, \citenamefont {Schuetz}, \citenamefont {Lukin}, \citenamefont {Loncar},\ and\ \citenamefont {Rabl}}]{Lemonde2018PhononWaveguides}%
  \BibitemOpen
  \bibfield  {author} {\bibinfo {author} {\bibfnamefont {M.~A.}\ \bibnamefont {Lemonde}}, \bibinfo {author} {\bibfnamefont {S.}~\bibnamefont {Meesala}}, \bibinfo {author} {\bibfnamefont {A.}~\bibnamefont {Sipahigil}}, \bibinfo {author} {\bibfnamefont {M.~J.}\ \bibnamefont {Schuetz}}, \bibinfo {author} {\bibfnamefont {M.~D.}\ \bibnamefont {Lukin}}, \bibinfo {author} {\bibfnamefont {M.}~\bibnamefont {Loncar}},\ and\ \bibinfo {author} {\bibfnamefont {P.}~\bibnamefont {Rabl}},\ }\bibfield  {title} {\bibinfo {title} {{Phonon Networks with Silicon-Vacancy Centers in Diamond Waveguides}},\ }\href {https://doi.org/10.1103/PHYSREVLETT.120.213603/FIGURES/4/MEDIUM} {\bibfield  {journal} {\bibinfo  {journal} {Physical Review Letters}\ }\textbf {\bibinfo {volume} {120}},\ \bibinfo {pages} {213603} (\bibinfo {year} {2018})}\BibitemShut {NoStop}%
\bibitem [{\citenamefont {Satzinger}\ \emph {et~al.}(2018)\citenamefont {Satzinger}, \citenamefont {Zhong}, \citenamefont {Chang}, \citenamefont {Peairs}, \citenamefont {Bienfait}, \citenamefont {Chou}, \citenamefont {Cleland}, \citenamefont {Conner}, \citenamefont {{Dumur}}, \citenamefont {Grebel}, \citenamefont {Gutierrez}, \citenamefont {November}, \citenamefont {Povey}, \citenamefont {Whiteley}, \citenamefont {Awschalom}, \citenamefont {Schuster},\ and\ \citenamefont {Cleland}}]{Satzinger2018QuantumPhonons}%
  \BibitemOpen
  \bibfield  {author} {\bibinfo {author} {\bibfnamefont {K.~J.}\ \bibnamefont {Satzinger}}, \bibinfo {author} {\bibfnamefont {Y.~P.}\ \bibnamefont {Zhong}}, \bibinfo {author} {\bibfnamefont {H.~S.}\ \bibnamefont {Chang}}, \bibinfo {author} {\bibfnamefont {G.~A.}\ \bibnamefont {Peairs}}, \bibinfo {author} {\bibfnamefont {A.}~\bibnamefont {Bienfait}}, \bibinfo {author} {\bibfnamefont {M.~H.}\ \bibnamefont {Chou}}, \bibinfo {author} {\bibfnamefont {A.~Y.}\ \bibnamefont {Cleland}}, \bibinfo {author} {\bibfnamefont {C.~R.}\ \bibnamefont {Conner}}, \bibinfo {author} {\bibnamefont {{Dumur}}}, \bibinfo {author} {\bibfnamefont {J.}~\bibnamefont {Grebel}}, \bibinfo {author} {\bibfnamefont {I.}~\bibnamefont {Gutierrez}}, \bibinfo {author} {\bibfnamefont {B.~H.}\ \bibnamefont {November}}, \bibinfo {author} {\bibfnamefont {R.~G.}\ \bibnamefont {Povey}}, \bibinfo {author} {\bibfnamefont {S.~J.}\ \bibnamefont {Whiteley}}, \bibinfo {author} {\bibfnamefont {D.~D.}\ \bibnamefont {Awschalom}}, \bibinfo {author} {\bibfnamefont
  {D.~I.}\ \bibnamefont {Schuster}},\ and\ \bibinfo {author} {\bibfnamefont {A.~N.}\ \bibnamefont {Cleland}},\ }\bibfield  {title} {\bibinfo {title} {{Quantum control of surface acoustic-wave phonons}},\ }\href {https://doi.org/10.1038/s41586-018-0719-5} {\bibfield  {journal} {\bibinfo  {journal} {Nature 2018 563:7733}\ }\textbf {\bibinfo {volume} {563}},\ \bibinfo {pages} {661} (\bibinfo {year} {2018})}\BibitemShut {NoStop}%
\bibitem [{\citenamefont {Cleland}\ \emph {et~al.}(2019)\citenamefont {Cleland}, \citenamefont {Pechal}, \citenamefont {Stas}, \citenamefont {Sarabalis}, \citenamefont {Wollack},\ and\ \citenamefont {Safavi-Naeini}}]{Cleland2019MechanicalMachines}%
  \BibitemOpen
  \bibfield  {author} {\bibinfo {author} {\bibfnamefont {A.~Y.}\ \bibnamefont {Cleland}}, \bibinfo {author} {\bibfnamefont {M.}~\bibnamefont {Pechal}}, \bibinfo {author} {\bibfnamefont {P.~J.~C.}\ \bibnamefont {Stas}}, \bibinfo {author} {\bibfnamefont {C.~J.}\ \bibnamefont {Sarabalis}}, \bibinfo {author} {\bibfnamefont {E.~A.}\ \bibnamefont {Wollack}},\ and\ \bibinfo {author} {\bibfnamefont {A.~H.}\ \bibnamefont {Safavi-Naeini}},\ }\bibfield  {title} {\bibinfo {title} {{Mechanical Purcell filters for microwave quantum machines}},\ }\href {https://doi.org/10.1063/1.5111151/13163607/263504{\_}1{\_}ACCEPTED{\_}MANUSCRIPT.PDF} {\bibfield  {journal} {\bibinfo  {journal} {Applied Physics Letters}\ }\textbf {\bibinfo {volume} {115}},\ \bibinfo {pages} {263504} (\bibinfo {year} {2019})}\BibitemShut {NoStop}%
\bibitem [{\citenamefont {Bienfait}\ \emph {et~al.}(2019)\citenamefont {Bienfait}, \citenamefont {Satzinger}, \citenamefont {Zhong}, \citenamefont {Chang}, \citenamefont {Chou}, \citenamefont {Conner}, \citenamefont {{Dumur}}, \citenamefont {Grebel}, \citenamefont {Peairs}, \citenamefont {Povey},\ and\ \citenamefont {Cleland}}]{Bienfait2019Phonon-mediatedEntanglement}%
  \BibitemOpen
  \bibfield  {author} {\bibinfo {author} {\bibfnamefont {A.}~\bibnamefont {Bienfait}}, \bibinfo {author} {\bibfnamefont {K.~J.}\ \bibnamefont {Satzinger}}, \bibinfo {author} {\bibfnamefont {Y.~P.}\ \bibnamefont {Zhong}}, \bibinfo {author} {\bibfnamefont {H.~S.}\ \bibnamefont {Chang}}, \bibinfo {author} {\bibfnamefont {M.~H.}\ \bibnamefont {Chou}}, \bibinfo {author} {\bibfnamefont {C.~R.}\ \bibnamefont {Conner}}, \bibinfo {author} {\bibnamefont {{Dumur}}}, \bibinfo {author} {\bibfnamefont {J.}~\bibnamefont {Grebel}}, \bibinfo {author} {\bibfnamefont {G.~A.}\ \bibnamefont {Peairs}}, \bibinfo {author} {\bibfnamefont {R.~G.}\ \bibnamefont {Povey}},\ and\ \bibinfo {author} {\bibfnamefont {A.~N.}\ \bibnamefont {Cleland}},\ }\bibfield  {title} {\bibinfo {title} {{Phonon-mediated quantum state transfer and remote qubit entanglement}},\ }\href {https://doi.org/10.1126/SCIENCE.AAW8415/SUPPL{\_}FILE/AAW8415{\_}BIENFAIT{\_}SM.PDF} {\bibfield  {journal} {\bibinfo  {journal} {Science}\ }\textbf {\bibinfo {volume} {364}},\
  \bibinfo {pages} {368} (\bibinfo {year} {2019})}\BibitemShut {NoStop}%
\bibitem [{\citenamefont {Pant}\ \emph {et~al.}(2011)\citenamefont {Pant}, \citenamefont {Poulton}, \citenamefont {Choi}, \citenamefont {Mcfarlane}, \citenamefont {Hile}, \citenamefont {Li}, \citenamefont {Thevenaz}, \citenamefont {Luther-Davies}, \citenamefont {Madden}, \citenamefont {Eggleton}, \citenamefont {Stenner}, \citenamefont {Neifeld}, \citenamefont {Gauthier}, \citenamefont {Boyd}, \citenamefont {Berger}, \citenamefont {Bourderionnet}, \citenamefont {Dolfi}, \citenamefont {Zhang}, \citenamefont {Yan}, \citenamefont {Yang}, \citenamefont {Fazal}, \citenamefont {Willner}, \citenamefont {Pelusi}, \citenamefont {Fu},\ and\ \citenamefont {Eggleton}}]{Pant2011On-chipScattering}%
  \BibitemOpen
  \bibfield  {author} {\bibinfo {author} {\bibfnamefont {R.}~\bibnamefont {Pant}}, \bibinfo {author} {\bibfnamefont {C.~G.}\ \bibnamefont {Poulton}}, \bibinfo {author} {\bibfnamefont {D.-Y.}\ \bibnamefont {Choi}}, \bibinfo {author} {\bibfnamefont {H.}~\bibnamefont {Mcfarlane}}, \bibinfo {author} {\bibfnamefont {S.}~\bibnamefont {Hile}}, \bibinfo {author} {\bibfnamefont {E.}~\bibnamefont {Li}}, \bibinfo {author} {\bibfnamefont {L.}~\bibnamefont {Thevenaz}}, \bibinfo {author} {\bibfnamefont {B.}~\bibnamefont {Luther-Davies}}, \bibinfo {author} {\bibfnamefont {S.~J.}\ \bibnamefont {Madden}}, \bibinfo {author} {\bibfnamefont {B.~J.}\ \bibnamefont {Eggleton}}, \bibinfo {author} {\bibfnamefont {M.~D.}\ \bibnamefont {Stenner}}, \bibinfo {author} {\bibfnamefont {M.~A.}\ \bibnamefont {Neifeld}}, \bibinfo {author} {\bibfnamefont {D.~J.}\ \bibnamefont {Gauthier}}, \bibinfo {author} {\bibfnamefont {R.~W.}\ \bibnamefont {Boyd}}, \bibinfo {author} {\bibfnamefont {P.}~\bibnamefont {Berger}}, \bibinfo {author} {\bibfnamefont
  {J.}~\bibnamefont {Bourderionnet}}, \bibinfo {author} {\bibfnamefont {D.}~\bibnamefont {Dolfi}}, \bibinfo {author} {\bibfnamefont {B.}~\bibnamefont {Zhang}}, \bibinfo {author} {\bibfnamefont {L.~S.}\ \bibnamefont {Yan}}, \bibinfo {author} {\bibfnamefont {J.~Y.}\ \bibnamefont {Yang}}, \bibinfo {author} {\bibfnamefont {I.}~\bibnamefont {Fazal}}, \bibinfo {author} {\bibfnamefont {A.~E.}\ \bibnamefont {Willner}}, \bibinfo {author} {\bibfnamefont {M.~D.}\ \bibnamefont {Pelusi}}, \bibinfo {author} {\bibfnamefont {A.}~\bibnamefont {Fu}},\ and\ \bibinfo {author} {\bibfnamefont {B.~J.}\ \bibnamefont {Eggleton}},\ }\bibfield  {title} {\bibinfo {title} {{On-chip stimulated Brillouin scattering}},\ }\href {https://doi.org/10.1364/OE.19.008285} {\bibfield  {journal} {\bibinfo  {journal} {Optics Express, Vol. 19, Issue 9, pp. 8285-8290}\ }\textbf {\bibinfo {volume} {19}},\ \bibinfo {pages} {8285} (\bibinfo {year} {2011})}\BibitemShut {NoStop}%
\bibitem [{\citenamefont {Fu}\ \emph {et~al.}(2019)\citenamefont {Fu}, \citenamefont {Shen}, \citenamefont {Xu}, \citenamefont {Zou}, \citenamefont {Cheng}, \citenamefont {Han},\ and\ \citenamefont {Tang}}]{Fu2019PhononicPhonons}%
  \BibitemOpen
  \bibfield  {author} {\bibinfo {author} {\bibfnamefont {W.}~\bibnamefont {Fu}}, \bibinfo {author} {\bibfnamefont {Z.}~\bibnamefont {Shen}}, \bibinfo {author} {\bibfnamefont {Y.}~\bibnamefont {Xu}}, \bibinfo {author} {\bibfnamefont {C.~L.}\ \bibnamefont {Zou}}, \bibinfo {author} {\bibfnamefont {R.}~\bibnamefont {Cheng}}, \bibinfo {author} {\bibfnamefont {X.}~\bibnamefont {Han}},\ and\ \bibinfo {author} {\bibfnamefont {H.~X.}\ \bibnamefont {Tang}},\ }\bibfield  {title} {\bibinfo {title} {{Phononic integrated circuitry and spin–orbit interaction of phonons}},\ }\href {https://doi.org/10.1038/s41467-019-10852-3} {\bibfield  {journal} {\bibinfo  {journal} {Nature Communications 2019 10:1}\ }\textbf {\bibinfo {volume} {10}},\ \bibinfo {pages} {1} (\bibinfo {year} {2019})}\BibitemShut {NoStop}%
\bibitem [{\citenamefont {Wang}\ \emph {et~al.}(2020)\citenamefont {Wang}, \citenamefont {Shen}, \citenamefont {Zou}, \citenamefont {Fu}, \citenamefont {Shen},\ and\ \citenamefont {Tang}}]{Wang2020High-acoustic-index-contrastModeling}%
  \BibitemOpen
  \bibfield  {author} {\bibinfo {author} {\bibfnamefont {W.}~\bibnamefont {Wang}}, \bibinfo {author} {\bibfnamefont {M.}~\bibnamefont {Shen}}, \bibinfo {author} {\bibfnamefont {C.~L.}\ \bibnamefont {Zou}}, \bibinfo {author} {\bibfnamefont {W.}~\bibnamefont {Fu}}, \bibinfo {author} {\bibfnamefont {Z.}~\bibnamefont {Shen}},\ and\ \bibinfo {author} {\bibfnamefont {H.~X.}\ \bibnamefont {Tang}},\ }\bibfield  {title} {\bibinfo {title} {{High-acoustic-index-contrast phononic circuits: Numerical modeling}},\ }\bibfield  {journal} {\bibinfo  {journal} {Journal of Applied Physics}\ }\textbf {\bibinfo {volume} {128}},\ \href {https://doi.org/10.1063/5.0019584/15256079/184503{\_}1{\_}ACCEPTED{\_}MANUSCRIPT.PDF} {10.1063/5.0019584/15256079/184503{\_}1{\_}ACCEPTED{\_}MANUSCRIPT.PDF} (\bibinfo {year} {2020})\BibitemShut {NoStop}%
\bibitem [{\citenamefont {Mayor}\ \emph {et~al.}(2021)\citenamefont {Mayor}, \citenamefont {Jiang}, \citenamefont {Sarabalis}, \citenamefont {McKenna}, \citenamefont {Witmer},\ and\ \citenamefont {Safavi-Naeini}}]{Mayor2021GigahertzSapphire}%
  \BibitemOpen
  \bibfield  {author} {\bibinfo {author} {\bibfnamefont {F.~M.}\ \bibnamefont {Mayor}}, \bibinfo {author} {\bibfnamefont {W.}~\bibnamefont {Jiang}}, \bibinfo {author} {\bibfnamefont {C.~J.}\ \bibnamefont {Sarabalis}}, \bibinfo {author} {\bibfnamefont {T.~P.}\ \bibnamefont {McKenna}}, \bibinfo {author} {\bibfnamefont {J.~D.}\ \bibnamefont {Witmer}},\ and\ \bibinfo {author} {\bibfnamefont {A.~H.}\ \bibnamefont {Safavi-Naeini}},\ }\bibfield  {title} {\bibinfo {title} {{Gigahertz Phononic Integrated Circuits on Thin-Film Lithium Niobate on Sapphire}},\ }\href {https://doi.org/10.1103/PHYSREVAPPLIED.15.014039/FIGURES/12/MEDIUM} {\bibfield  {journal} {\bibinfo  {journal} {Physical Review Applied}\ }\textbf {\bibinfo {volume} {15}},\ \bibinfo {pages} {014039} (\bibinfo {year} {2021})}\BibitemShut {NoStop}%
\bibitem [{\citenamefont {Bicer}\ \emph {et~al.}(2022)\citenamefont {Bicer}, \citenamefont {Valle}, \citenamefont {Brown}, \citenamefont {Kuball},\ and\ \citenamefont {Balram}}]{Bicer2022GalliumDevices}%
  \BibitemOpen
  \bibfield  {author} {\bibinfo {author} {\bibfnamefont {M.}~\bibnamefont {Bicer}}, \bibinfo {author} {\bibfnamefont {S.}~\bibnamefont {Valle}}, \bibinfo {author} {\bibfnamefont {J.}~\bibnamefont {Brown}}, \bibinfo {author} {\bibfnamefont {M.}~\bibnamefont {Kuball}},\ and\ \bibinfo {author} {\bibfnamefont {K.~C.}\ \bibnamefont {Balram}},\ }\bibfield  {title} {\bibinfo {title} {{Gallium nitride phononic integrated circuits platform for GHz frequency acoustic wave devices}},\ }\href {https://doi.org/10.1063/5.0082467/2833881} {\bibfield  {journal} {\bibinfo  {journal} {Applied Physics Letters}\ }\textbf {\bibinfo {volume} {120}},\ \bibinfo {pages} {243502} (\bibinfo {year} {2022})}\BibitemShut {NoStop}%
\bibitem [{\citenamefont {Guida}\ and\ \citenamefont {Ghosh}(2024)}]{Guida2024DesignCircuits}%
  \BibitemOpen
  \bibfield  {author} {\bibinfo {author} {\bibfnamefont {J.}~\bibnamefont {Guida}}\ and\ \bibinfo {author} {\bibfnamefont {S.}~\bibnamefont {Ghosh}},\ }\bibfield  {title} {\bibinfo {title} {{Design and Analysis of Guided Surface Acoustic Waves in ScAlN on Sapphire for Phononic Integrated Circuits}},\ }\bibfield  {journal} {\bibinfo  {journal} {IEEE Transactions on Ultrasonics, Ferroelectrics, and Frequency Control}\ }\href {https://doi.org/10.1109/TUFFC.2024.3484181} {10.1109/TUFFC.2024.3484181} (\bibinfo {year} {2024})\BibitemShut {NoStop}%
\bibitem [{\citenamefont {Hatanaka}\ \emph {et~al.}(2014)\citenamefont {Hatanaka}, \citenamefont {Mahboob}, \citenamefont {Onomitsu},\ and\ \citenamefont {Yamaguchi}}]{Hatanaka2014PhononCircuits}%
  \BibitemOpen
  \bibfield  {author} {\bibinfo {author} {\bibfnamefont {D.}~\bibnamefont {Hatanaka}}, \bibinfo {author} {\bibfnamefont {I.}~\bibnamefont {Mahboob}}, \bibinfo {author} {\bibfnamefont {K.}~\bibnamefont {Onomitsu}},\ and\ \bibinfo {author} {\bibfnamefont {H.}~\bibnamefont {Yamaguchi}},\ }\bibfield  {title} {\bibinfo {title} {{Phonon waveguides for electromechanical circuits}},\ }\href {https://doi.org/10.1038/nnano.2014.107} {\bibfield  {journal} {\bibinfo  {journal} {Nature Nanotechnology 2014 9:7}\ }\textbf {\bibinfo {volume} {9}},\ \bibinfo {pages} {520} (\bibinfo {year} {2014})}\BibitemShut {NoStop}%
\bibitem [{\citenamefont {Van~Laer}\ \emph {et~al.}(2018)\citenamefont {Van~Laer}, \citenamefont {Patel}, \citenamefont {McKenna}, \citenamefont {Witmer},\ and\ \citenamefont {Safavi-Naeini}}]{VanLaer2018ElectricalCircuit}%
  \BibitemOpen
  \bibfield  {author} {\bibinfo {author} {\bibfnamefont {R.}~\bibnamefont {Van~Laer}}, \bibinfo {author} {\bibfnamefont {R.~N.}\ \bibnamefont {Patel}}, \bibinfo {author} {\bibfnamefont {T.~P.}\ \bibnamefont {McKenna}}, \bibinfo {author} {\bibfnamefont {J.~D.}\ \bibnamefont {Witmer}},\ and\ \bibinfo {author} {\bibfnamefont {A.~H.}\ \bibnamefont {Safavi-Naeini}},\ }\bibfield  {title} {\bibinfo {title} {{Electrical driving of X-band mechanical waves in a silicon photonic circuit}},\ }\href {https://doi.org/10.1063/1.5042428/19738160/086102{\_}1{\_}ACCEPTED{\_}MANUSCRIPT.PDF} {\bibfield  {journal} {\bibinfo  {journal} {APL Photonics}\ }\textbf {\bibinfo {volume} {3}},\ \bibinfo {pages} {86102} (\bibinfo {year} {2018})}\BibitemShut {NoStop}%
\bibitem [{\citenamefont {Liu}\ \emph {et~al.}(2019)\citenamefont {Liu}, \citenamefont {Li},\ and\ \citenamefont {Li}}]{Liu2019ElectromechanicalWaveguides}%
  \BibitemOpen
  \bibfield  {author} {\bibinfo {author} {\bibfnamefont {Q.}~\bibnamefont {Liu}}, \bibinfo {author} {\bibfnamefont {H.}~\bibnamefont {Li}},\ and\ \bibinfo {author} {\bibfnamefont {M.}~\bibnamefont {Li}},\ }\bibfield  {title} {\bibinfo {title} {{Electromechanical Brillouin scattering in integrated optomechanical waveguides}},\ }\href {https://doi.org/10.1364/OPTICA.6.000778} {\bibfield  {journal} {\bibinfo  {journal} {Optica, Vol. 6, Issue 6, pp. 778-785}\ }\textbf {\bibinfo {volume} {6}},\ \bibinfo {pages} {778} (\bibinfo {year} {2019})}\BibitemShut {NoStop}%
\bibitem [{\citenamefont {Romero}\ \emph {et~al.}(2019)\citenamefont {Romero}, \citenamefont {Kalra}, \citenamefont {Mauranyapin}, \citenamefont {Baker}, \citenamefont {Meng},\ and\ \citenamefont {Bowen}}]{Romero2019PropagationWaveguide}%
  \BibitemOpen
  \bibfield  {author} {\bibinfo {author} {\bibfnamefont {E.}~\bibnamefont {Romero}}, \bibinfo {author} {\bibfnamefont {R.}~\bibnamefont {Kalra}}, \bibinfo {author} {\bibfnamefont {N.~P.}\ \bibnamefont {Mauranyapin}}, \bibinfo {author} {\bibfnamefont {C.~G.}\ \bibnamefont {Baker}}, \bibinfo {author} {\bibfnamefont {C.}~\bibnamefont {Meng}},\ and\ \bibinfo {author} {\bibfnamefont {W.~P.}\ \bibnamefont {Bowen}},\ }\bibfield  {title} {\bibinfo {title} {{Propagation and Imaging of Mechanical Waves in a Highly Stressed Single-Mode Acoustic Waveguide}},\ }\href {https://doi.org/10.1103/PHYSREVAPPLIED.11.064035/FIGURES/5/MEDIUM} {\bibfield  {journal} {\bibinfo  {journal} {Physical Review Applied}\ }\textbf {\bibinfo {volume} {11}},\ \bibinfo {pages} {064035} (\bibinfo {year} {2019})}\BibitemShut {NoStop}%
\bibitem [{\citenamefont {Sarabalis}\ \emph {et~al.}(2020)\citenamefont {Sarabalis}, \citenamefont {Dahmani}, \citenamefont {Cleland},\ and\ \citenamefont {Safavi-Naeini}}]{Sarabalis2020S-bandNiobate}%
  \BibitemOpen
  \bibfield  {author} {\bibinfo {author} {\bibfnamefont {C.~J.}\ \bibnamefont {Sarabalis}}, \bibinfo {author} {\bibfnamefont {Y.~D.}\ \bibnamefont {Dahmani}}, \bibinfo {author} {\bibfnamefont {A.~Y.}\ \bibnamefont {Cleland}},\ and\ \bibinfo {author} {\bibfnamefont {A.~H.}\ \bibnamefont {Safavi-Naeini}},\ }\bibfield  {title} {\bibinfo {title} {{S-band delay lines in suspended lithium niobate}},\ }\href {https://doi.org/10.1063/1.5126428/15241382/054501{\_}1{\_}ACCEPTED{\_}MANUSCRIPT.PDF} {\bibfield  {journal} {\bibinfo  {journal} {Journal of Applied Physics}\ }\textbf {\bibinfo {volume} {127}},\ \bibinfo {pages} {54501} (\bibinfo {year} {2020})}\BibitemShut {NoStop}%
\bibitem [{\citenamefont {Dahmani}\ \emph {et~al.}(2020)\citenamefont {Dahmani}, \citenamefont {Sarabalis}, \citenamefont {Jiang}, \citenamefont {Mayor},\ and\ \citenamefont {Safavi-Naeini}}]{Dahmani2020PiezoelectricWaveguide}%
  \BibitemOpen
  \bibfield  {author} {\bibinfo {author} {\bibfnamefont {Y.~D.}\ \bibnamefont {Dahmani}}, \bibinfo {author} {\bibfnamefont {C.~J.}\ \bibnamefont {Sarabalis}}, \bibinfo {author} {\bibfnamefont {W.}~\bibnamefont {Jiang}}, \bibinfo {author} {\bibfnamefont {F.~M.}\ \bibnamefont {Mayor}},\ and\ \bibinfo {author} {\bibfnamefont {A.~H.}\ \bibnamefont {Safavi-Naeini}},\ }\bibfield  {title} {\bibinfo {title} {{Piezoelectric Transduction of a Wavelength-Scale Mechanical Waveguide}},\ }\href {https://doi.org/10.1103/PHYSREVAPPLIED.13.024069/FIGURES/8/MEDIUM} {\bibfield  {journal} {\bibinfo  {journal} {Physical Review Applied}\ }\textbf {\bibinfo {volume} {13}},\ \bibinfo {pages} {024069} (\bibinfo {year} {2020})}\BibitemShut {NoStop}%
\bibitem [{\citenamefont {Lu}\ \emph {et~al.}(2020)\citenamefont {Lu}, \citenamefont {Yang}, \citenamefont {Li}, \citenamefont {Manzaneque},\ and\ \citenamefont {Gong}}]{Lu2020GHzLines}%
  \BibitemOpen
  \bibfield  {author} {\bibinfo {author} {\bibfnamefont {R.}~\bibnamefont {Lu}}, \bibinfo {author} {\bibfnamefont {Y.}~\bibnamefont {Yang}}, \bibinfo {author} {\bibfnamefont {M.~H.}\ \bibnamefont {Li}}, \bibinfo {author} {\bibfnamefont {T.}~\bibnamefont {Manzaneque}},\ and\ \bibinfo {author} {\bibfnamefont {S.}~\bibnamefont {Gong}},\ }\bibfield  {title} {\bibinfo {title} {{GHz Broadband SH0 Mode Lithium Niobate Acoustic Delay Lines}},\ }\href {https://doi.org/10.1109/TUFFC.2019.2943355} {\bibfield  {journal} {\bibinfo  {journal} {IEEE Transactions on Ultrasonics, Ferroelectrics, and Frequency Control}\ }\textbf {\bibinfo {volume} {67}},\ \bibinfo {pages} {402} (\bibinfo {year} {2020})}\BibitemShut {NoStop}%
\bibitem [{\citenamefont {Kurosu}\ \emph {et~al.}(2020)\citenamefont {Kurosu}, \citenamefont {Hatanaka},\ and\ \citenamefont {Yamaguchi}}]{Kurosu2020MechanicalWaveguide}%
  \BibitemOpen
  \bibfield  {author} {\bibinfo {author} {\bibfnamefont {M.}~\bibnamefont {Kurosu}}, \bibinfo {author} {\bibfnamefont {D.}~\bibnamefont {Hatanaka}},\ and\ \bibinfo {author} {\bibfnamefont {H.}~\bibnamefont {Yamaguchi}},\ }\bibfield  {title} {\bibinfo {title} {{Mechanical Kerr Nonlinearity of Wave Propagation in an On-Chip Nanoelectromechanical Waveguide}},\ }\href {https://doi.org/10.1103/PHYSREVAPPLIED.13.014056/FIGURES/9/MEDIUM} {\bibfield  {journal} {\bibinfo  {journal} {Physical Review Applied}\ }\textbf {\bibinfo {volume} {13}},\ \bibinfo {pages} {014056} (\bibinfo {year} {2020})}\BibitemShut {NoStop}%
\bibitem [{\citenamefont {Ding}\ \emph {et~al.}(2024)\citenamefont {Ding}, \citenamefont {Pingault}, \citenamefont {Shao}, \citenamefont {Sinclair}, \citenamefont {Machielse}, \citenamefont {Chia}, \citenamefont {Maity},\ and\ \citenamefont {Lon{\v{c}}ar}}]{Ding2024IntegratedDiamond}%
  \BibitemOpen
  \bibfield  {author} {\bibinfo {author} {\bibfnamefont {S.~W.}\ \bibnamefont {Ding}}, \bibinfo {author} {\bibfnamefont {B.}~\bibnamefont {Pingault}}, \bibinfo {author} {\bibfnamefont {L.}~\bibnamefont {Shao}}, \bibinfo {author} {\bibfnamefont {N.}~\bibnamefont {Sinclair}}, \bibinfo {author} {\bibfnamefont {B.}~\bibnamefont {Machielse}}, \bibinfo {author} {\bibfnamefont {C.}~\bibnamefont {Chia}}, \bibinfo {author} {\bibfnamefont {S.}~\bibnamefont {Maity}},\ and\ \bibinfo {author} {\bibfnamefont {M.}~\bibnamefont {Lon{\v{c}}ar}},\ }\bibfield  {title} {\bibinfo {title} {{Integrated phononic waveguides in diamond}},\ }\href {https://doi.org/10.1103/PHYSREVAPPLIED.21.014034/FIGURES/8/MEDIUM} {\bibfield  {journal} {\bibinfo  {journal} {Physical Review Applied}\ }\textbf {\bibinfo {volume} {21}},\ \bibinfo {pages} {014034} (\bibinfo {year} {2024})}\BibitemShut {NoStop}%
\bibitem [{\citenamefont {Feng}\ \emph {et~al.}(2023)\citenamefont {Feng}, \citenamefont {Liu}, \citenamefont {Xi}, \citenamefont {Wang},\ and\ \citenamefont {Sun}}]{Feng2023GigahertzWaveguides}%
  \BibitemOpen
  \bibfield  {author} {\bibinfo {author} {\bibfnamefont {Z.}~\bibnamefont {Feng}}, \bibinfo {author} {\bibfnamefont {Y.}~\bibnamefont {Liu}}, \bibinfo {author} {\bibfnamefont {X.}~\bibnamefont {Xi}}, \bibinfo {author} {\bibfnamefont {L.}~\bibnamefont {Wang}},\ and\ \bibinfo {author} {\bibfnamefont {X.}~\bibnamefont {Sun}},\ }\bibfield  {title} {\bibinfo {title} {{Gigahertz Phononic Integrated Circuits Based on Overlay Slot Waveguides}},\ }\href {https://doi.org/10.1103/PHYSREVAPPLIED.19.064076/FIGURES/6/MEDIUM} {\bibfield  {journal} {\bibinfo  {journal} {Physical Review Applied}\ }\textbf {\bibinfo {volume} {19}},\ \bibinfo {pages} {064076} (\bibinfo {year} {2023})}\BibitemShut {NoStop}%
\bibitem [{\citenamefont {Sun}\ and\ \citenamefont {Wu}(2006)}]{Sun2006PropagationMethod}%
  \BibitemOpen
  \bibfield  {author} {\bibinfo {author} {\bibfnamefont {J.~H.}\ \bibnamefont {Sun}}\ and\ \bibinfo {author} {\bibfnamefont {T.~T.}\ \bibnamefont {Wu}},\ }\bibfield  {title} {\bibinfo {title} {{Propagation of surface acoustic waves through sharply bent two-dimensional phononic crystal waveguides using a finite-difference time-domain method}},\ }\href {https://doi.org/10.1103/PhysRevB.74.174305} {\bibfield  {journal} {\bibinfo  {journal} {Physical Review B}\ }\textbf {\bibinfo {volume} {74}},\ \bibinfo {pages} {174305} (\bibinfo {year} {2006})}\BibitemShut {NoStop}%
\bibitem [{\citenamefont {Vasseur}\ \emph {et~al.}(2007)\citenamefont {Vasseur}, \citenamefont {Hladky-Hennion}, \citenamefont {Djafari-Rouhani}, \citenamefont {Duval}, \citenamefont {Dubus}, \citenamefont {Pennec},\ and\ \citenamefont {Deymier}}]{Vasseur2007WaveguidingPlates}%
  \BibitemOpen
  \bibfield  {author} {\bibinfo {author} {\bibfnamefont {J.~O.}\ \bibnamefont {Vasseur}}, \bibinfo {author} {\bibfnamefont {A.~C.}\ \bibnamefont {Hladky-Hennion}}, \bibinfo {author} {\bibfnamefont {B.}~\bibnamefont {Djafari-Rouhani}}, \bibinfo {author} {\bibfnamefont {F.}~\bibnamefont {Duval}}, \bibinfo {author} {\bibfnamefont {B.}~\bibnamefont {Dubus}}, \bibinfo {author} {\bibfnamefont {Y.}~\bibnamefont {Pennec}},\ and\ \bibinfo {author} {\bibfnamefont {P.~A.}\ \bibnamefont {Deymier}},\ }\bibfield  {title} {\bibinfo {title} {{Waveguiding in two-dimensional piezoelectric phononic crystal plates}},\ }\href {https://doi.org/10.1063/1.2740352/987979} {\bibfield  {journal} {\bibinfo  {journal} {Journal of Applied Physics}\ }\textbf {\bibinfo {volume} {101}},\ \bibinfo {pages} {114904} (\bibinfo {year} {2007})}\BibitemShut {NoStop}%
\bibitem [{\citenamefont {Mohammadi}\ \emph {et~al.}(2009)\citenamefont {Mohammadi}, \citenamefont {Eftekhar}, \citenamefont {Hunt},\ and\ \citenamefont {Adibi}}]{Mohammadi2009High-QSlab}%
  \BibitemOpen
  \bibfield  {author} {\bibinfo {author} {\bibfnamefont {S.}~\bibnamefont {Mohammadi}}, \bibinfo {author} {\bibfnamefont {A.~A.}\ \bibnamefont {Eftekhar}}, \bibinfo {author} {\bibfnamefont {W.~D.}\ \bibnamefont {Hunt}},\ and\ \bibinfo {author} {\bibfnamefont {A.}~\bibnamefont {Adibi}},\ }\bibfield  {title} {\bibinfo {title} {{High-Q micromechanical resonators in a two-dimensional phononic crystal slab}},\ }\href {https://doi.org/10.1063/1.3078284/337626} {\bibfield  {journal} {\bibinfo  {journal} {Applied Physics Letters}\ }\textbf {\bibinfo {volume} {94}},\ \bibinfo {pages} {51906} (\bibinfo {year} {2009})}\BibitemShut {NoStop}%
\bibitem [{\citenamefont {Khelif}\ \emph {et~al.}(2010)\citenamefont {Khelif}, \citenamefont {Mohammadi}, \citenamefont {Eftekhar}, \citenamefont {Adibi},\ and\ \citenamefont {Aoubiza}}]{Khelif2010AcousticSlabs}%
  \BibitemOpen
  \bibfield  {author} {\bibinfo {author} {\bibfnamefont {A.}~\bibnamefont {Khelif}}, \bibinfo {author} {\bibfnamefont {S.}~\bibnamefont {Mohammadi}}, \bibinfo {author} {\bibfnamefont {A.~A.}\ \bibnamefont {Eftekhar}}, \bibinfo {author} {\bibfnamefont {A.}~\bibnamefont {Adibi}},\ and\ \bibinfo {author} {\bibfnamefont {B.}~\bibnamefont {Aoubiza}},\ }\bibfield  {title} {\bibinfo {title} {{Acoustic confinement and waveguiding with a line-defect structure in phononic crystal slabs}},\ }\href {https://doi.org/10.1063/1.3500226/938171} {\bibfield  {journal} {\bibinfo  {journal} {Journal of Applied Physics}\ }\textbf {\bibinfo {volume} {108}},\ \bibinfo {pages} {84515} (\bibinfo {year} {2010})}\BibitemShut {NoStop}%
\bibitem [{\citenamefont {Balram}\ \emph {et~al.}(2016)\citenamefont {Balram}, \citenamefont {Davan{\c{c}}o}, \citenamefont {Song},\ and\ \citenamefont {Srinivasan}}]{Balram2016CoherentCircuits}%
  \BibitemOpen
  \bibfield  {author} {\bibinfo {author} {\bibfnamefont {K.~C.}\ \bibnamefont {Balram}}, \bibinfo {author} {\bibfnamefont {M.~I.}\ \bibnamefont {Davan{\c{c}}o}}, \bibinfo {author} {\bibfnamefont {J.~D.}\ \bibnamefont {Song}},\ and\ \bibinfo {author} {\bibfnamefont {K.}~\bibnamefont {Srinivasan}},\ }\bibfield  {title} {\bibinfo {title} {{Coherent coupling between radiofrequency, optical and acoustic waves in piezo-optomechanical circuits}},\ }\href {https://doi.org/10.1038/nphoton.2016.46} {\bibfield  {journal} {\bibinfo  {journal} {Nature Photonics 2016 10:5}\ }\textbf {\bibinfo {volume} {10}},\ \bibinfo {pages} {346} (\bibinfo {year} {2016})}\BibitemShut {NoStop}%
\bibitem [{\citenamefont {Patel}\ \emph {et~al.}(2018)\citenamefont {Patel}, \citenamefont {Wang}, \citenamefont {Jiang}, \citenamefont {Sarabalis}, \citenamefont {Hill},\ and\ \citenamefont {Safavi-Naeini}}]{Patel2018Single-ModeWire}%
  \BibitemOpen
  \bibfield  {author} {\bibinfo {author} {\bibfnamefont {R.~N.}\ \bibnamefont {Patel}}, \bibinfo {author} {\bibfnamefont {Z.}~\bibnamefont {Wang}}, \bibinfo {author} {\bibfnamefont {W.}~\bibnamefont {Jiang}}, \bibinfo {author} {\bibfnamefont {C.~J.}\ \bibnamefont {Sarabalis}}, \bibinfo {author} {\bibfnamefont {J.~T.}\ \bibnamefont {Hill}},\ and\ \bibinfo {author} {\bibfnamefont {A.~H.}\ \bibnamefont {Safavi-Naeini}},\ }\bibfield  {title} {\bibinfo {title} {{Single-Mode Phononic Wire}},\ }\href {https://doi.org/10.1103/PHYSREVLETT.121.040501/FIGURES/4/MEDIUM} {\bibfield  {journal} {\bibinfo  {journal} {Physical Review Letters}\ }\textbf {\bibinfo {volume} {121}},\ \bibinfo {pages} {040501} (\bibinfo {year} {2018})}\BibitemShut {NoStop}%
\bibitem [{\citenamefont {Pourabolghasem}\ \emph {et~al.}(2018)\citenamefont {Pourabolghasem}, \citenamefont {Dehghannasiri}, \citenamefont {Eftekhar},\ and\ \citenamefont {Adibi}}]{Pourabolghasem2018WaveguidingSlabs}%
  \BibitemOpen
  \bibfield  {author} {\bibinfo {author} {\bibfnamefont {R.}~\bibnamefont {Pourabolghasem}}, \bibinfo {author} {\bibfnamefont {R.}~\bibnamefont {Dehghannasiri}}, \bibinfo {author} {\bibfnamefont {A.~A.}\ \bibnamefont {Eftekhar}},\ and\ \bibinfo {author} {\bibfnamefont {A.}~\bibnamefont {Adibi}},\ }\bibfield  {title} {\bibinfo {title} {{Waveguiding Effect in the Gigahertz Frequency Range in Pillar-based Phononic-Crystal Slabs}},\ }\href {https://doi.org/10.1103/PHYSREVAPPLIED.9.014013/FIGURES/11/MEDIUM} {\bibfield  {journal} {\bibinfo  {journal} {Physical Review Applied}\ }\textbf {\bibinfo {volume} {9}},\ \bibinfo {pages} {014013} (\bibinfo {year} {2018})}\BibitemShut {NoStop}%
\bibitem [{\citenamefont {Wang}\ \emph {et~al.}(2019)\citenamefont {Wang}, \citenamefont {Lee}, \citenamefont {Zheng}, \citenamefont {Xie},\ and\ \citenamefont {Feng}}]{Wang2019HexagonalWaveguides}%
  \BibitemOpen
  \bibfield  {author} {\bibinfo {author} {\bibfnamefont {Y.}~\bibnamefont {Wang}}, \bibinfo {author} {\bibfnamefont {J.}~\bibnamefont {Lee}}, \bibinfo {author} {\bibfnamefont {X.~Q.}\ \bibnamefont {Zheng}}, \bibinfo {author} {\bibfnamefont {Y.}~\bibnamefont {Xie}},\ and\ \bibinfo {author} {\bibfnamefont {P.~X.}\ \bibnamefont {Feng}},\ }\bibfield  {title} {\bibinfo {title} {{Hexagonal Boron Nitride Phononic Crystal Waveguides}},\ }\href {https://doi.org/10.1021/ACSPHOTONICS.9B01094/ASSET/IMAGES/LARGE/PH9B01094{\_}0006.JPEG} {\bibfield  {journal} {\bibinfo  {journal} {ACS Photonics}\ }\textbf {\bibinfo {volume} {6}},\ \bibinfo {pages} {3225} (\bibinfo {year} {2019})}\BibitemShut {NoStop}%
\bibitem [{\citenamefont {Chen}\ \emph {et~al.}(2023)\citenamefont {Chen}, \citenamefont {Liu}, \citenamefont {Li}, \citenamefont {Song}, \citenamefont {Wang}, \citenamefont {Wang}, \citenamefont {Wang},\ and\ \citenamefont {Zhu}}]{Chen2023Scandium-DopedApplications}%
  \BibitemOpen
  \bibfield  {author} {\bibinfo {author} {\bibfnamefont {L.}~\bibnamefont {Chen}}, \bibinfo {author} {\bibfnamefont {C.}~\bibnamefont {Liu}}, \bibinfo {author} {\bibfnamefont {M.}~\bibnamefont {Li}}, \bibinfo {author} {\bibfnamefont {W.}~\bibnamefont {Song}}, \bibinfo {author} {\bibfnamefont {W.}~\bibnamefont {Wang}}, \bibinfo {author} {\bibfnamefont {Z.}~\bibnamefont {Wang}}, \bibinfo {author} {\bibfnamefont {N.}~\bibnamefont {Wang}},\ and\ \bibinfo {author} {\bibfnamefont {Y.}~\bibnamefont {Zhu}},\ }\bibfield  {title} {\bibinfo {title} {{Scandium-Doped Aluminum Nitride for Acoustic Wave Resonators, Filters, and Ferroelectric Memory Applications}},\ }\href {https://doi.org/10.1021/ACSAELM.2C01409/ASSET/IMAGES/LARGE/EL2C01409{\_}0005.JPEG} {\bibfield  {journal} {\bibinfo  {journal} {ACS Applied Electronic Materials}\ }\textbf {\bibinfo {volume} {5}},\ \bibinfo {pages} {612} (\bibinfo {year} {2023})}\BibitemShut {NoStop}%
\bibitem [{\citenamefont {Hackett}\ \emph {et~al.}(2023{\natexlab{b}})\citenamefont {Hackett}, \citenamefont {Miller}, \citenamefont {Beaucejour}, \citenamefont {Nordquist}, \citenamefont {Taylor}, \citenamefont {Santillan}, \citenamefont {Olsson},\ and\ \citenamefont {Eichenfield}}]{Hackett2023AluminumFrequencies}%
  \BibitemOpen
  \bibfield  {author} {\bibinfo {author} {\bibfnamefont {L.}~\bibnamefont {Hackett}}, \bibinfo {author} {\bibfnamefont {M.}~\bibnamefont {Miller}}, \bibinfo {author} {\bibfnamefont {R.}~\bibnamefont {Beaucejour}}, \bibinfo {author} {\bibfnamefont {C.~M.}\ \bibnamefont {Nordquist}}, \bibinfo {author} {\bibfnamefont {J.~C.}\ \bibnamefont {Taylor}}, \bibinfo {author} {\bibfnamefont {S.}~\bibnamefont {Santillan}}, \bibinfo {author} {\bibfnamefont {R.~H.}\ \bibnamefont {Olsson}},\ and\ \bibinfo {author} {\bibfnamefont {M.}~\bibnamefont {Eichenfield}},\ }\bibfield  {title} {\bibinfo {title} {{Aluminum scandium nitride films for piezoelectric transduction into silicon at gigahertz frequencies}},\ }\href {https://doi.org/10.1063/5.0151434/2906810} {\bibfield  {journal} {\bibinfo  {journal} {Applied Physics Letters}\ }\textbf {\bibinfo {volume} {123}},\ \bibinfo {pages} {73502} (\bibinfo {year} {2023}{\natexlab{b}})}\BibitemShut {NoStop}%
\bibitem [{\citenamefont {Du}\ \emph {et~al.}(2024)\citenamefont {Du}, \citenamefont {Sharma}, \citenamefont {Tang}, \citenamefont {Leblanc}, \citenamefont {Jariwala},\ and\ \citenamefont {Olsson}}]{Du2024NearSiC}%
  \BibitemOpen
  \bibfield  {author} {\bibinfo {author} {\bibfnamefont {X.}~\bibnamefont {Du}}, \bibinfo {author} {\bibfnamefont {N.}~\bibnamefont {Sharma}}, \bibinfo {author} {\bibfnamefont {Z.}~\bibnamefont {Tang}}, \bibinfo {author} {\bibfnamefont {C.}~\bibnamefont {Leblanc}}, \bibinfo {author} {\bibfnamefont {D.}~\bibnamefont {Jariwala}},\ and\ \bibinfo {author} {\bibfnamefont {R.~H.}\ \bibnamefont {Olsson}},\ }\bibfield  {title} {\bibinfo {title} {{Near 6-GHz Sezawa Mode Surface Acoustic Wave Resonators Using AlScN on SiC}},\ }\href {https://doi.org/10.1109/JMEMS.2024.3430984} {\bibfield  {journal} {\bibinfo  {journal} {Journal of Microelectromechanical Systems}\ ,\ \bibinfo {pages} {1}} (\bibinfo {year} {2024})}\BibitemShut {NoStop}%
\bibitem [{\citenamefont {Luke}\ \emph {et~al.}(2020)\citenamefont {Luke}, \citenamefont {Kharel}, \citenamefont {Reimer}, \citenamefont {He}, \citenamefont {Loncar}, \citenamefont {Loncar},\ and\ \citenamefont {Zhang}}]{Luke2020Wafer-scaleCircuits}%
  \BibitemOpen
  \bibfield  {author} {\bibinfo {author} {\bibfnamefont {K.}~\bibnamefont {Luke}}, \bibinfo {author} {\bibfnamefont {P.}~\bibnamefont {Kharel}}, \bibinfo {author} {\bibfnamefont {C.}~\bibnamefont {Reimer}}, \bibinfo {author} {\bibfnamefont {L.}~\bibnamefont {He}}, \bibinfo {author} {\bibfnamefont {M.}~\bibnamefont {Loncar}}, \bibinfo {author} {\bibfnamefont {M.}~\bibnamefont {Loncar}},\ and\ \bibinfo {author} {\bibfnamefont {M.}~\bibnamefont {Zhang}},\ }\bibfield  {title} {\bibinfo {title} {{Wafer-scale low-loss lithium niobate photonic integrated circuits}},\ }\href {https://doi.org/10.1364/OE.401959} {\bibfield  {journal} {\bibinfo  {journal} {Optics Express, Vol. 28, Issue 17, pp. 24452-24458}\ }\textbf {\bibinfo {volume} {28}},\ \bibinfo {pages} {24452} (\bibinfo {year} {2020})}\BibitemShut {NoStop}%
\bibitem [{\citenamefont {Castelletto}\ \emph {et~al.}(2020)\citenamefont {Castelletto}, \citenamefont {Boretti}, \citenamefont {Phys~Photonics}, \citenamefont {Moody}, \citenamefont {Sorger},\ and\ \citenamefont {Blumenthal}}]{Castelletto2020SiliconApplications}%
  \BibitemOpen
  \bibfield  {author} {\bibinfo {author} {\bibfnamefont {S.}~\bibnamefont {Castelletto}}, \bibinfo {author} {\bibfnamefont {A.}~\bibnamefont {Boretti}}, \bibinfo {author} {\bibfnamefont {J.}~\bibnamefont {Phys~Photonics}}, \bibinfo {author} {\bibfnamefont {G.}~\bibnamefont {Moody}}, \bibinfo {author} {\bibfnamefont {V.~J.}\ \bibnamefont {Sorger}},\ and\ \bibinfo {author} {\bibfnamefont {D.~J.}\ \bibnamefont {Blumenthal}},\ }\bibfield  {title} {\bibinfo {title} {{Silicon carbide color centers for quantum applications}},\ }\href {https://doi.org/10.1088/2515-7647/AB77A2} {\bibfield  {journal} {\bibinfo  {journal} {Journal of Physics: Photonics}\ }\textbf {\bibinfo {volume} {2}},\ \bibinfo {pages} {022001} (\bibinfo {year} {2020})}\BibitemShut {NoStop}%
\bibitem [{\citenamefont {Buijtendorp}\ \emph {et~al.}(2022)\citenamefont {Buijtendorp}, \citenamefont {Vollebregt}, \citenamefont {Karatsu}, \citenamefont {Thoen}, \citenamefont {Murugesan}, \citenamefont {Kouwenhoven}, \citenamefont {H{\"{a}}hnle}, \citenamefont {Baselmans},\ and\ \citenamefont {Endo}}]{Buijtendorp2022HydrogenatedCircuits}%
  \BibitemOpen
  \bibfield  {author} {\bibinfo {author} {\bibfnamefont {B.~T.}\ \bibnamefont {Buijtendorp}}, \bibinfo {author} {\bibfnamefont {S.}~\bibnamefont {Vollebregt}}, \bibinfo {author} {\bibfnamefont {K.}~\bibnamefont {Karatsu}}, \bibinfo {author} {\bibfnamefont {D.~J.}\ \bibnamefont {Thoen}}, \bibinfo {author} {\bibfnamefont {V.}~\bibnamefont {Murugesan}}, \bibinfo {author} {\bibfnamefont {K.}~\bibnamefont {Kouwenhoven}}, \bibinfo {author} {\bibfnamefont {S.}~\bibnamefont {H{\"{a}}hnle}}, \bibinfo {author} {\bibfnamefont {J.~J.~A.}\ \bibnamefont {Baselmans}},\ and\ \bibinfo {author} {\bibfnamefont {A.}~\bibnamefont {Endo}},\ }\bibfield  {title} {\bibinfo {title} {{Hydrogenated Amorphous Silicon Carbide: A Low-Loss Deposited Dielectric for Microwave to Submillimeter-Wave Superconducting Circuits}},\ }\href {https://doi.org/10.1103/PhysRevApplied.18.064003} {\bibfield  {journal} {\bibinfo  {journal} {Physical Review Applied}\ }\textbf {\bibinfo {volume} {10}},\ \bibinfo {pages} {64003} (\bibinfo {year}
  {2022})}\BibitemShut {NoStop}%
\end{thebibliography}%

\end{document}